\documentclass[journal ]{new-aiaa}
\usepackage[utf8]{inputenc}
\usepackage{textcomp}
\usepackage{graphicx}
\usepackage{amsmath}
\usepackage[version=4]{mhchem}
\usepackage{siunitx}
\usepackage{longtable,tabularx}
\usepackage{hyperref}
\usepackage{latexsym}
\usepackage{multirow}
\usepackage{makecell}
\usepackage{graphicx}
\usepackage{float}
\usepackage{subfigure}
\usepackage{subcaption}

\title{Time-Varying Directional State Transition Tensor for Orbit Uncertainty Propagation}

\author{Xingyu Zhou\footnote{Ph.D. Candidate, School of Aerospace Engineering; \url{zhouxingyu@bit.edu.cn}.}}
\affil{Beijing Institute of Technology, 100081 Beijing, People’s Republic of China}
\author{Roberto Armellin\footnote{Professor, Te Pūnaha Ātea - Space Institute, 20 Symonds Street, Auckland Central; \url{roberto.armellin@auckland.ac.nz} (Member AIAA).}}
\affil{University of Auckland, Auckland 1010, New Zealand}
\author{Dong Qiao\footnote{Professor, School of Aerospace Engineering; \url{qiaodong@bit.edu.cn} (Corresponding Author).} and Xiangyu Li\footnote{Associate Professor, School of Aerospace Engineering; \url{lixiangy@bit.edu.cn}.}}
\affil{Beijing Institute of Technology, 100081 Beijing, People’s Republic of China}

\begin{document}

\maketitle

\begin{abstract}
The directional state transition tensor (DSTT) reduces the complexity of state transition tensor (STT) by aligning the STT terms in sensitive directions only, which provides comparable accuracy in orbital uncertainty propagation. The DSTT assumes the sensitive directions to be constant during the integration and only works at a predefined epoch. This paper proposes a time-varying STT (TDSTT) to improve the DSTT. The proposed TDSTT computes the sensitive directions with time; thereby, it can perform uncertainty propagation analysis at any point instead of only a predefined epoch as the DSTT does. First, the derivatives of the sensitive directions are derived. Then, the differential equations for the high-order TDSTTs are derived and simplified using the orthogonality of sensitive directions. Next, complexity analysis is implemented to show the advantages of the proposed TDSTT over the STT. Finally, the TDSTT is applied to solve orbital uncertainty propagation problems in highly nonlinear three-body systems. Numerical results show that the proposed TDSTT can yield nearly the same level of accuracy as the STT and DSTT. It is approximately 94\% faster than the STT and has hundreds of improvements in speed over the DSTT when one wants to investigate the evolutions of orbital uncertainties.
\end{abstract}

\section{Introduction}
\lettrine{O}{rbital} uncertainty propagation is a key technique in astrodynamics \cite{Luo2017PAS}. Namely, it is used to predict the uncertainties in orbital states and is widely used in space tasks such as orbit determination \cite{Gong2023Astro, Zhou2024JGCD, Fossa2024}, orbital sensitivity analysis \cite{Jenson2022JGCD, Jenson2024}, conjunction assessment and collision avoidance \cite{Armellin2021, Uriot2022, Khatri2023}, and orbital guidance and control \cite{Boone2020JGCD}. In addition, using orbital uncertainty propagation techniques to implement robust trajectory optimization has become a tendency in recent mission design processes \cite{Jenson2021, Greco2022, Yang2023SST}.

In theory, the exact solution for orbital uncertainty propagation can only be obtained by solving the Fokker-Planck equation (FPE) \cite{Luo2017PAS}. However, the FPE is a high-dimensional nonlinear partial differential equation (PDE). For a general orbital uncertainty propagation problem, the dimension is six, and a high-order dimension is required if additional parameters (\emph{e.g.}, maneuver parameters for maneuver reconstruction \cite{Zhou2024JGCD, Zhou2022Aerospace, JiaRichards2023, Pirovano2024, Zhang2024SST}, and spherical harmonic coefficients for gravitational field reconstruction \cite{Yin2024Astro}) are considered. Thus, the PDE of the FPE is usually difficult to solve as it doesn’t have analytical solutions. Monte Carlo (MC) simulation is another way to solve the orbital uncertainty propagation problem. It involves repeatedly sampling random values for uncertain parameters (usually the initial orbital states) from their respective probability distributions and then propagating these values through the orbital dynamics model to obtain a distribution of potential orbital states \cite{Luo2017PAS}. By propagating a large number of orbital trajectories, the MC method provides a statistical representation of the orbital state distributions. Although accurate, the MC suffers from a high computational burden as it requires repeated propagation \cite{Qiao2023ASR}.

A variety of methods have been developed to approximate the solution to improve the computational efficiency of orbital uncertainty propagation, which can be broadly divided into two categories: linear and nonlinear methods \cite{Luo2017PAS}. Classic linear methods include the covariance analysis describing function technique (CADET) \cite{Gelb1973} and the state transformation matrix (STM) \cite{Qiao2023Astro, Gim2003}. These linear methods suffer from drawbacks related to poor accuracy as they linearize the nonlinear dynamics. They may be capable of short-period uncertainty propagation in a dynamic environment whose nonlinearity is not that high; however, they are insufficient for operations in highly nonlinear dynamics such as the orbits in a multi-body system \cite{Park2006} (\emph{e.g.}, Earth-Moon, Sun-Earth, and Sun-Jupiter systems) or around an irregular asteroid \cite{Feng2019}.

To address the linearization deficiencies, high-order nonlinear methods have been developed, including unscented transformation (UT) \cite{Julier2000, Xiong2006}, conjugate UT (CUT) \cite{Adurthi2015, Adurthi2018, Nanda2018}, and high-order Taylor polynomials \cite{Park2006, Armellin2010}.
The UT and CUT belong to the sigma-point methods, which deterministically select the samples and propagate them to approximate the final distributions \cite{Julier2000, Adurthi2015}. The UT and CUT require significantly fewer samples than the MC.
High-order Taylor polynomial approximation of the solution of ordinary differential equations (ODEs) are mainly obtained via variational approach, producing the so-called state transition tensor (STT) \cite{Park2006, Park2007, Majji2008}, or by implementing the algebra of Taylor polynomials, like in differential algebra (DA) \cite{BerzModernMM, Armellin2010, Feng2019, Shu2022JGCD} and jet transport (JT) \cite{Palau2015}. The variational approach requires the computation of complex partial derivatives of the dynamics and the integration of a larger set of ODEs. DA and JT instead integrate the original set of ODEs but require custom integration schemes to handle the Taylor polynomials data type \cite{Pavanello2024, Palau2015}.
Compared with the sigma-point methods, the high-order Taylor polynomials have two significant advantages. Firstly, the high-order Taylor polynomials can be repeatedly employed when the initial state distributions change, while samples should be re-generated for UT and CUT. Secondly, the high-order Taylor polynomials can produce the statistical representation of the orbital state distribution (distribution-to-distribution) and predict the deviation states from a point-to-point way, which is extremely useful when applied to orbital guidance and control \cite{Yang2019Astro, Boone2020JGCD}.

One of the drawbacks of the above high-order Taylor polynomial methods lies in that the number of terms increases exponentially as the order increases, leading to an exponential growth of the computational burden. Several efforts have been made to reduce computational costs. Roa and Park proposed a reduced STT (RSTT), which assumes that the two-body gravitational effects dominate the dynamics. The RSTT reduces the high-order terms of the STT by considering only the secular terms \cite{Roa2021}. Zhou \emph{et al.} employed an analytical STT to predict the state deviations under the two-body dynamics and trained a deep neural network (DNN) to compensate for the effects due to the perturbations. Its applications are limited as the DNN must be pre-trained \cite{Zhou2024IEEE}. A recent significant progress in this field is the directional STT (DSTT) proposed by Boone and McMahon \cite{Boone2022JGCD}. The DSTT computes the sensitive direction of the maximum stretch of the Cauchy-Green tensor (CGT) at a predefined epoch and aligns the STT terms in the (sensitive) direction of the eigenvector associated with the largest eigenvalue of the CGT. In this way, the DSTT requires significantly fewer terms to predict the uncertainties. Boone and McMahon then applied the DSTT to estimate the orbit in the Earth-Moon three-body system \cite{Boodram2022, Boone2022, Boone2024}. Qiao \emph{et al.} employed the DSTT to accelerate the configuration uncertainty propagation of a space-based gravitational-wave observatory \cite{QIAO2024CJA}. In the above work (\emph{i.e.}, \cite{Boone2022JGCD, Boodram2022, Boone2022, Boone2024, QIAO2024CJA}), the sensitive directions are fixed during the integration of the DSTT terms; thus, the DSTT terms cannot be extracted along the trajectory evolution, as the sensitive directions are determined at a specific epoch. This feature might not be a limitation when solving an orbit estimation problem as this process only requires the orbital uncertainty predictions at predefined measurement epochs (\emph{i.e.}, the final epoch of one measurement interval). However, the evolution of the orbital uncertainties is important for applications such as conjunction assessment and collision avoidance. If one has to perform an uncertainty analysis at different epochs, a new DSTT must be integrated as the sensitive directions change, leading to additional computational expense. These points yield the motivations of this work.

This paper develops a time-varying DSTT (referred to as TDSTT) to enhance STT and DSTT methods. Notably, the proposed approach is based on the DSTT framework, but it can avoid the additional computational cost when applied to analyze orbital uncertainties on a dense time grid. First, the TDSTT framework is established, in which the nominal orbits, the high-order Taylor series expansion, and the eigenvalue-eigenvector pairs are integrated simultaneously. The eigenvalue-eigenvector pairs vary over time, and the direction changes during the integration. In this way, one can extract the results during the integration of TDSTT to predict or analyze the orbital uncertainties at any historical point. Then, the derivatives of the TDSTTs are derived. The computational complexity for calculating the TDSTT derivatives is significantly reduced by using the eigenvectors’ orthogonality properties. The overall algorithm complexity analysis is carried out for both the TDSTT and its competitive methods. It will be shown that the algorithm complexity of the TDSTT grows polynomially instead of exponentially. Finally, the TDSTT method is applied to solve the uncertainty propagation in highly nonlinear cases: a temporary capture (TC) orbit in the Sun-Jupiter three-body system and a near-rectilinear halo orbit (NRHO) in the Earth-Moon three-body system.

The remainder of this paper is organized as follows. Section~\ref{Sec:Review of State Transition Tensor} briefly reviews the STT and DSTT techniques. The derivatives of the eigenvalue-eigenvector pairs are presented in Sec.~\ref{Sec:Derivatives of Eigenvalues and Eigenvectors}. Section~\ref{Sec:Time-varying Directional State Transition Tensor} begins with introducing the basic idea of the proposed TDSTT, and then the detailed derivation process of the TDSTT’s derivatives is shown in Sec.~\ref{Sec:Derivatives of TDSTT}. Overall procedure is summarized in Sec.~\ref{Sec:Overall Procedure} and the algorithm complexity is analyzed in Sec.~\ref{Sec:Complexity analysis}. Numerical simulations are presented in Sec.~\ref{Sec:Numerical Examples}, and conclusions are given in Sec.~\ref{Sec:Conclusion}.

\section{Review of State Transition Tensor} \label{Sec:Review of State Transition Tensor}

\subsection{Full State Transition Tensor} \label{Sec:Full State Transition Tensor}
Consider a general format of a nonlinear orbital dynamics model governed by an ODE as
\begin{equation} \label{eq:dynamics}
    \boldsymbol{\dot x} = \boldsymbol{f}(\boldsymbol{x},t) \,,
\end{equation}
where $\boldsymbol{x} \in \mathbb{R}^{n}$ is the orbit state (vector) at a given epoch $t$ ($n$ is the dimension of the state and $n=6$ for orbital state), and $\boldsymbol{f}:{\mathbb{R}^n} \mapsto {\mathbb{R}^n}$ represents the nonlinear orbital dynamics. Let $\boldsymbol{x}_{0} \in \mathbb{R}^{n}$ be the orbit state at an initial epoch $t_{0}$; then, the orbit state $\boldsymbol{x}$ at any given epoch $t$ can be obtained by solving the ODE in Eq.~\eqref{eq:dynamics}, written as
\begin{equation} \label{eq:state solution}
    \boldsymbol{x} = \boldsymbol{F}(t;{\boldsymbol{x}_0},{t_0}) = \int_{{t_0}}^t {\boldsymbol{f}(\boldsymbol{x},\tau ){\rm{d}}\tau }  + {\boldsymbol{x}_0} \,.
\end{equation}

Let $\delta \boldsymbol{x}_{0} \in \mathbb{R}^{n}$ be the initial state deviation (\emph{i.e.}, the deviation between the nominal and neighboring orbits at the initial epoch $t_{0}$); then, the state deviation at epoch $t$ is defined as $\delta \boldsymbol{x} = \boldsymbol{F}(t;{\boldsymbol{x}_0} + \delta {\boldsymbol{x}_0},{t_0}) - \boldsymbol{F}(t;{\boldsymbol{x}_0},{t_0})$. According to Ref.~\cite{Park2006}, a Taylor expansion series can be used to analytically predict $\delta \boldsymbol{x}$, given as
\begin{equation} \label{eq:Taylor expansion series}
    \delta {x^i} \approx \sum\limits_{p = 1}^P {\frac{1}{{p!}}\phi _{({t_0},t)}^{i,{k_1} \cdots {k_p}}\delta x_0^{{k_1}} \cdots \delta x_0^{{k_p}}}  = \phi _{({t_0},t)}^{i,{k_1}}\delta x_0^{{k_1}} + \frac{1}{2}\phi _{({t_0},t)}^{i,{k_1}{k_2}}\delta x_0^{{k_1}}\delta x_0^{{k_2}} + \frac{1}{6}\phi _{({t_0},t)}^{i,{k_1}{k_2}{k_3}}\delta x_0^{{k_1}}\delta x_0^{{k_2}}\delta x_0^{{k_3}} +  \cdots \,,
\end{equation}
where $\delta {x^i}$ represents the \emph{i}-th elements ($i \in \{ 1, \cdots ,n\}$) of the state deviation $\delta \boldsymbol{x}$, $\delta x_0^{{k_p}}$ represents the $k_{p}$-th variable of the initial state deviation $\delta \boldsymbol{x}_{0}$, $P$ is the highest order of the Taylor expansion series (a user-defined parameter), and $\phi _{({t_0},t)}^{i,{k_1} \cdots {k_p}}$ is the \emph{p}-th order (full) STT, defined by
\begin{equation} \label{eq:STT}
    \phi _{({t_0},t)}^{i,{k_1} \cdots {k_p}} = \frac{{{\partial ^p}{x^i}}}{{\partial x_0^{{k_1}} \cdots \partial x_0^{{k_p}}}} \,.
\end{equation}

Generally, the first-order STT is also called the STM (\emph{i.e.}, $\phi _{({t_0},t)}^{i,{k_1}}$ is the STM element at the \emph{i}-th row, $k_{1}$-th column). Note that the Einstein summation notation is employed in Eq.~\eqref{eq:Taylor expansion series} (and throughout this paper). The dummy variables (superscripts) in Eq.~\eqref{eq:Taylor expansion series} (\emph{e.g.}, $k_{1}$, $k_{2}$, and $k_{p}$) increment from 1 to $n$ (recall that $n=6$ for the orbital state) during the summation process. Taking the first- and second-order STTs as examples, one has $\phi _{({t_0},t)}^{i,{k_1}}\delta x_0^{{k_1}} = \sum\limits_{{k_1} = 1}^n {\phi _{({t_0},t)}^{i,{k_1}}\delta x_0^{{k_1}}}$ and $\phi _{({t_0},t)}^{i,{k_1}{k_2}}\delta x_0^{{k_1}}\delta x_0^{{k_2}} = \sum\limits_{{k_1} = 1}^n {\sum\limits_{{k_2} = 1}^n {\phi _{({t_0},t)}^{i,{k_1}{k_2}}\delta x_0^{{k_1}}\delta x_0^{{k_2}}} }$.

The STTs can be obtained by integrating the corresponding ODEs along the nominal orbit. The following equations show the ODEs for the first- through fourth-order STTs:
\begin{equation} \label{eq:STT1 ODE}
    \dot \phi _{({t_0},t)}^{i,{k_1}} = {A^{i,\alpha }}\phi _{({t_0},t)}^{\alpha ,{k_1}} \,,
\end{equation}
\begin{equation} \label{eq:STT2 ODE}
    \dot \phi _{({t_0},t)}^{i,{k_1}{k_2}} = {A^{i,\alpha }}\phi _{({t_0},t)}^{\alpha ,{k_1}{k_2}} + {A^{i,\alpha \beta }}\phi _{({t_0},t)}^{\alpha ,{k_1}}\phi _{({t_0},t)}^{\beta ,{k_2}} \,,
\end{equation}
\begin{equation} \label{eq:STT3 ODE}
    \begin{aligned}
        \dot \phi _{({t_0},t)}^{i,{k_1}{k_2}{k_3}} &= {A^{i,\alpha }}\phi _{({t_0},t)}^{\alpha ,{k_1}{k_2}{k_3}} + {A^{i,\alpha \beta }}\left[ {\phi _{({t_0},t)}^{\alpha ,{k_1}}\phi _{({t_0},t)}^{\beta ,{k_2}{k_3}} + \phi _{({t_0},t)}^{\alpha ,{k_1}{k_2}}\phi _{({t_0},t)}^{\beta ,{k_3}} + \phi _{({t_0},t)}^{\alpha ,{k_1}{k_3}}\phi _{({t_0},t)}^{\beta ,{k_2}}} \right]\\
        &+ {A^{i,\alpha \beta \gamma }}\phi _{({t_0},t)}^{\alpha ,{k_1}}\phi _{({t_0},t)}^{\beta ,{k_2}}\phi _{({t_0},t)}^{\gamma ,{k_3}}
    \end{aligned} \,,
\end{equation}
\begin{equation} \label{eq:STT4 ODE}
    \begin{aligned}
        \dot \phi _{({t_0},t)}^{i,{k_1}{k_2}{k_3}{k_4}} &= {A^{i,\alpha }}\phi _{({t_0},t)}^{\alpha ,{k_1}{k_2}{k_3}{k_4}} + {A^{i,\alpha \beta }}\left[ {\phi _{({t_0},t)}^{\alpha ,{k_1}{k_2}{k_3}}\phi _{({t_0},t)}^{\beta ,{k_4}} + \phi _{({t_0},t)}^{\alpha ,{k_1}{k_2}{k_4}}\phi _{({t_0},t)}^{\beta ,{k_3}} + \phi _{({t_0},t)}^{\alpha ,{k_1}{k_3}{k_4}}\phi _{({t_0},t)}^{\beta ,{k_2}}} \right.\\
        & + \left. {\phi _{({t_0},t)}^{\alpha ,{k_1}{k_2}}\phi _{({t_0},t)}^{\beta ,{k_3}{k_4}} + \phi _{({t_0},t)}^{\alpha ,{k_1}{k_3}}\phi _{({t_0},t)}^{\beta ,{k_2}{k_4}} + \phi _{({t_0},t)}^{\alpha ,{k_1}{k_4}}\phi _{({t_0},t)}^{\beta ,{k_2}{k_3}} + \phi _{({t_0},t)}^{\alpha ,{k_1}}\phi _{({t_0},t)}^{\beta ,{k_2}{k_3}{k_4}}} \right]\\
        & + {A^{i,\alpha \beta \gamma }}\left[ {\phi _{({t_0},t)}^{\alpha ,{k_1}{k_2}}\phi _{({t_0},t)}^{\beta ,{k_3}}\phi _{({t_0},t)}^{\gamma ,{k_4}} + \phi _{({t_0},t)}^{\alpha ,{k_1}{k_3}}\phi _{({t_0},t)}^{\beta ,{k_2}}\phi _{({t_0},t)}^{\gamma ,{k_4}} + \phi _{({t_0},t)}^{\alpha ,{k_1}{k_4}}\phi _{({t_0},t)}^{\beta ,{k_2}}\phi _{({t_0},t)}^{\gamma ,{k_3}}} \right.\\
        & + \left. {\phi _{({t_0},t)}^{\alpha ,{k_1}}\phi _{({t_0},t)}^{\beta ,{k_2}{k_3}}\phi _{({t_0},t)}^{\gamma ,{k_4}} + \phi _{({t_0},t)}^{\alpha ,{k_1}}\phi _{({t_0},t)}^{\beta ,{k_2}{k_4}}\phi _{({t_0},t)}^{\gamma ,{k_3}} + \phi _{({t_0},t)}^{\alpha ,{k_1}}\phi _{({t_0},t)}^{\beta ,{k_2}}\phi _{({t_0},t)}^{\gamma ,{k_3}{k_4}}} \right]\\
        & + {A^{i,\alpha \beta \gamma \delta }}\phi _{({t_0},t)}^{\alpha ,{k_1}}\phi _{({t_0},t)}^{\beta ,{k_2}}\phi _{({t_0},t)}^{\gamma ,{k_3}}\phi _{({t_0},t)}^{\delta ,{k_4}}
    \end{aligned} \,,
\end{equation}
where $\alpha$, $\beta$, $\gamma$, and $\delta$ are dummy variables (\emph{i.e.}, they increment from 1 to $n$ during the summation process), and ${A^{i,{k_1} \cdots {k_p}}}$ is the $p$-th order local Jacobi element, which is defined as the derivative of the nonlinear dynamics (\emph{i.e.}, $\boldsymbol{f}(\boldsymbol{x},t)$) with respect to the state $\boldsymbol{x}$, formulated as
\begin{equation} \label{eq:A}
    {A^{i,{k_1} \cdots {k_p}}} = \frac{{{\partial ^p}{f^i}(\boldsymbol{x},t)}}{{\partial {x^{{k_1}}} \cdots \partial {x^{{k_p}}}}} \,,
\end{equation}
where ${f^i}(\boldsymbol{x},t)$ represents the $i$-th element of the vector $\boldsymbol{f}(\boldsymbol{x},t)$.

Notably, high-order STTs tend to be more precise than the STMs. However, the $p$-th order STT contains $n^{p+1}$ variables, and clearly, the computational burden of the STT grows exponentially as its order increases. Specifically, the first-order STT (STM) necessitates the integration of additional $n^2$ variables (besides the $n$ variables for the nominal orbit), and that amount increases to $n^3$ for the second-order STT. One of the most significant progress in improving the efficiency of high-order STTs is the DSTT, which will be briefly introduced in the next subsection.

\subsection{Directional State Transition Tensor} \label{Sec:Directional State Transition Tensor}
The DSTT technique reduces the computational cost of the high-order STTs by aligning the STTs with sensitive directions and ignoring the STT terms of the stable directions \cite{Boone2022JGCD}. Let $\boldsymbol{R} = {[{R^{ij}}]_{m \times n}} \in {\mathbb{R}^{m \times n}}$ ($m \le n$) be a linear transformation matrix, with $m$ being the number of sensitive directions and each row of $\boldsymbol{R}$ corresponding to a sensitive direction. Using the linear transformation matrix $\boldsymbol{R}$, a reduced state vector $\boldsymbol{y}$ can be obtained as
\begin{equation} \label{eq:y}
    \boldsymbol{y} = \boldsymbol{Rx} \in {\mathbb{R}^m} \,.
\end{equation}

According to Eq.~\eqref{eq:y}, one has ${{\partial {x^{{k_1}}}} \mathord{\left/{\vphantom {{\partial {x^{{k_1}}}} {\partial {y^{{p_1}}}}}}\right.
\kern-\nulldelimiterspace} {\partial {y^{{p_1}}}}} = {R^{{p_1},{k_1}}}$ \cite{Boone2022JGCD}, where ${R^{{p_1},{k_1}}}$ denotes the element at the $p_{1}$-th row, $k_{1}$-th column of the matrix $\boldsymbol{R}$. The DSTT is the derivative of the orbital state vector $\boldsymbol{x}$ with respect to the initial reduced state $\boldsymbol{y}_{0}\in {\mathbb{R}^m}$, defined as
\begin{equation} \label{eq:DSTT}
    \varphi _{({t_0},t)}^{i,{p_1} \cdots {p_p}} = \frac{{{\partial ^p}{x^i}}}{{\partial y_0^{{p_1}} \cdots \partial y_0^{{p_p}}}} = \frac{{{\partial ^p}{x^i}}}{{\partial x_0^{{k_1}} \cdots \partial x_0^{{k_p}}}}\frac{{\partial x_0^{{k_1}}}}{{\partial y_0^{{p_1}}}} \cdots \frac{{\partial x_0^{{k_p}}}}{{\partial y_0^{{p_p}}}} = \phi _{({t_0},t)}^{i,{k_1} \cdots {k_p}}{R^{{p_1},{k_1}}} \cdots {R^{{p_p},{k_p}}} \,,
\end{equation}
where $\varphi _{({t_0},t)}^{i,{p_1} \cdots {p_p}}$ denotes the $p$-th order DSTT. According to Eq.~\eqref{eq:DSTT}, the DSTTs up to the fourth order are expressed as
\begin{equation} \label{eq:DSTT1}
    \varphi _{({t_0},{t_k})}^{i,{p_1}} = \phi _{({t_0},{t_k})}^{i,{k_1}}{R^{{p_1},{k_1}}} \,,
\end{equation}
\begin{equation} \label{eq:DSTT2}
    \varphi _{({t_0},t)}^{i,{p_1}{p_2}} = \phi _{({t_0},t)}^{i,{k_1}{k_2}}{R^{{p_1},{k_1}}}{R^{{p_2},{k_2}}} \,,
\end{equation}
\begin{equation} \label{eq:DSTT3}
    \varphi _{({t_0},t)}^{i,{p_1}{p_2}{p_3}} = \phi _{({t_0},t)}^{i,{k_1}{k_2}{k_3}}{R^{{p_1},{k_1}}}{R^{{p_2},{k_2}}}{R^{{p_3},{k_3}}} \,,
\end{equation}
\begin{equation} \label{eq:DSTT4}
    \varphi _{({t_0},t)}^{i,{p_1}{p_2}{p_3}{p_4}} = \phi _{({t_0},t)}^{i,{k_1}{k_2}{k_3}{k_4}}{R^{{p_1},{k_1}}}{R^{{p_2},{k_2}}}{R^{{p_3},{k_3}}}{R^{{p_4},{k_4}}} \,.
\end{equation}

Once the linear transformation matrix $\boldsymbol{R}$ is determined, the derivatives of the DSTTs can be directly derived. The ODE for the first-order DSTT is derived as \cite{Boone2022JGCD}
\begin{equation} \label{eq:DSTT1 ODE}
    \dot \varphi _{({t_0},t)}^{i,{p_1}} = (\phi _{({t_0},{t_k})}^{i,{k_1}}{R^{{p_1},{k_1}}})' = \dot \phi _{({t_0},{t_k})}^{i,{k_1}}{R^{{p_1},{k_1}}} = {A^{i,\alpha }}\phi _{({t_0},{t_k})}^{\alpha ,{k_1}}{R^{{p_1},{k_1}}} = {A^{i,\alpha }}\varphi _{({t_0},t)}^{\alpha ,{p_1}} \,.
\end{equation}
Using a similar strategy, the ODEs for the second-, third-, and fourth-order DSTTs can be written as \cite{Boone2022JGCD}
\begin{equation} \label{eq:DSTT2 ODE}
    \dot \varphi _{({t_0},t)}^{i,{p_1}{p_2}} = {A^{i,\alpha }}\varphi _{({t_0},t)}^{\alpha ,{p_1}{p_2}} + {A^{i,\alpha \beta }}\varphi _{({t_0},t)}^{\alpha ,{p_1}}\varphi _{({t_0},t)}^{\beta ,{p_2}} \,,
\end{equation}
\begin{equation} \label{eq:DSTT3 ODE}
    \begin{aligned}
        \dot \varphi _{({t_0},t)}^{i,{p_1}{p_2}{p_3}} &= {A^{i,\alpha \beta }}\left[ {\varphi _{({t_0},t)}^{\alpha ,{p_1}}\varphi _{({t_0},t)}^{\beta ,{p_2}{p_3}} + \varphi _{({t_0},t)}^{\alpha ,{p_1}{p_2}}\varphi _{({t_0},t)}^{\beta ,{p_3}} + \varphi _{({t_0},t)}^{\alpha ,{p_1}{p_3}}\varphi _{({t_0},t)}^{\beta ,{p_2}}} \right]\\
        &+ {A^{i,\alpha }}\varphi _{({t_0},t)}^{\alpha ,{p_1}{p_2}{p_3}} + {A^{i,\alpha \beta \lambda }}\varphi _{({t_0},t)}^{\alpha ,{p_1}}\varphi _{({t_0},t)}^{\beta ,{p_2}}\varphi _{({t_0},t)}^{\lambda ,{p_3}}
    \end{aligned} \,,
\end{equation}
\begin{equation} \label{eq:DSTT4 ODE}
    \begin{aligned}
        \dot \phi _{({t_0},t)}^{i,{p_1}{p_2}{p_3}{p_4}} &= {A^{i,\alpha }}\phi _{({t_0},t)}^{\alpha ,{p_1}{p_2}{p_3}{p_4}} + {A^{i,\alpha \beta }}\left[ {\phi _{({t_0},t)}^{\alpha ,{p_1}{p_2}{p_3}}\phi _{({t_0},t)}^{\beta ,{p_4}} + \phi _{({t_0},t)}^{\alpha ,{p_1}{p_2}{p_4}}\phi _{({t_0},t)}^{\beta ,{p_3}} + \phi _{({t_0},t)}^{\alpha ,{p_1}{p_3}{p_4}}\phi _{({t_0},t)}^{\beta ,{p_2}}} \right.\\
        &+ \left. {\phi _{({t_0},t)}^{\alpha ,{p_1}{p_2}}\phi _{({t_0},t)}^{\beta ,{p_3}{p_4}} + \phi _{({t_0},t)}^{\alpha ,{p_1}{p_3}}\phi _{({t_0},t)}^{\beta ,{p_2}{p_4}} + \phi _{({t_0},t)}^{\alpha ,{p_1}{p_4}}\phi _{({t_0},t)}^{\beta ,{p_2}{p_3}} + \phi _{({t_0},t)}^{\alpha ,{p_1}}\phi _{({t_0},t)}^{\beta ,{p_2}{p_3}{p_4}}} \right]\\
        &+ {A^{i,\alpha \beta \lambda }}\left[ {\phi _{({t_0},t)}^{\alpha ,{p_1}{p_2}}\phi _{({t_0},t)}^{\beta ,{p_3}}\phi _{({t_0},t)}^{\lambda ,{p_4}} + \phi _{({t_0},t)}^{\alpha ,{p_1}{p_3}}\phi _{({t_0},t)}^{\beta ,{p_2}}\phi _{({t_0},t)}^{\lambda ,{p_4}} + \phi _{({t_0},t)}^{\alpha ,{p_1}d}\phi _{({t_0},t)}^{\beta ,{p_2}}\phi _{({t_0},t)}^{\lambda ,{p_3}}} \right.\\
        &+ \left. {\phi _{({t_0},t)}^{\alpha ,{p_1}}\phi _{({t_0},t)}^{\beta ,{p_2}{p_3}}\phi _{({t_0},t)}^{\lambda ,{p_4}} + \phi _{({t_0},t)}^{\alpha ,{p_1}}\phi _{({t_0},t)}^{\beta ,{p_2}{p_4}}\phi _{({t_0},t)}^{\lambda ,{p_3}} + \phi _{({t_0},t)}^{\alpha ,{p_1}}\phi _{({t_0},t)}^{\beta ,{p_2}}\phi _{({t_0},t)}^{\lambda ,{p_3}{p_4}}} \right]\\
        &+ {A^{i,\alpha \beta \lambda \delta }}\phi _{({t_0},t)}^{\alpha ,{p_1}}\phi _{({t_0},t)}^{\beta ,{p_2}}\phi _{({t_0},t)}^{\lambda ,{p_3}}\phi _{({t_0},t)}^{\delta ,{p_4}}
    \end{aligned} \,.
\end{equation}
Notice that the derivation process in Eqs.~\eqref{eq:DSTT1 ODE}-\eqref{eq:DSTT4 ODE} relies on the assumption that the linear transformation matrix $\boldsymbol{R}$ is constant (\emph{i.e.}, ${\dot R^{{p_1},{k_1}}} = {\dot R^{{p_p},{k_p}}} = 0$).

In Ref.~\cite{Boone2022JGCD}, the first-order STT (STM) is preserved, and the higher-order full STTs are approximated using the corresponding DSTT; thus, Eq.~\eqref{eq:Taylor expansion series} can be restated as
\begin{equation} \label{eq:Taylor DSTT}
    \delta {x^i} \approx \phi _{({t_0},t)}^{i,{k_1}}\delta x_0^{{k_1}} + \sum\limits_{p = 2}^P {\frac{1}{{p!}}\varphi _{({t_0},t)}^{i,{p_1} \cdots {p_p}}\delta y_0^{{p_1}} \cdots \delta y_0^{{p_p}}} \,.
\end{equation}

Compared with the full STT (as shown in Eq.~\eqref{eq:Taylor expansion series}), the improvement in computational burden using the DSTT (Eq.~\eqref{eq:Taylor DSTT} is significant. Numerical integration of $\sum\nolimits_{p = 0}^P {{n^{p + 1}}}$ variables is required if Eq.~\eqref{eq:Taylor expansion series} is employed to predict the state deviation $\delta \boldsymbol{x}$, and such a number reduces to $n + {n^2} + \sum\nolimits_{p = 2}^P {n{m^p}}$ if the second- and higher-order (full) STTs are replaced using DSTTs. To be more specific, for the case of $P=2$, $n=6$, and $m=1$ (\emph{i.e.}, using one sensitive direction), a total of 258 variables (6 variables for the nominal orbit, 36 variables for the STM, and 216 variables for the second-order STT) are integrated for the STT, while only 48 variables (additional 6 variables for the second-order DSTT) are required using the DSTT.

In Ref.~\cite{Boone2022JGCD}, the sensitive directions (\emph{i.e.}, the linear transformation matrix $\boldsymbol{R}$) are selected based on the CGT. To be convenient, let ${\boldsymbol{\Phi }_{({t_0},t)}} = {[\phi _{({t_0},t)}^{i,{k_1}}]_{n \times n}} \in {\mathbb{R}^{n \times n}}$ be the STM from initial epoch $t_{0}$ to epoch $t$. The CGT is defined as the STM multiplied by its transpose, expressed as
\begin{equation} \label{eq:CGT}
    \boldsymbol{C} = \mathbf{\Phi }_{({t_0},t)}^T{\mathbf{\Phi }_{({t_0},t)}} \in {\mathbb{R}^{n \times n}} \,,
\end{equation}
where $\boldsymbol{C}$ is the CGT from initial epoch $t_{0}$ to $t$ (the subscript $\left ( t_{0}, t \right )$ is not shown for the sake of simplicity), which is a symmetric matrix of size $n \times n$. Using the relation $\delta \boldsymbol{x} = {\mathbf{\Phi }_{({t_0},t)}}\delta {\boldsymbol{x}_0}$, one has
\begin{equation} \label{eq:Norm of dx}
    \begin{aligned}
        {\left\| {\delta \boldsymbol{x}} \right\|^2} &= \delta {\boldsymbol{x}^T}\delta \boldsymbol{x} = {({\mathbf{\Phi }_{({t_0},t)}}\delta {\boldsymbol{x}_0})^T}{\mathbf{\Phi }_{({t_0},t)}}\delta {\boldsymbol{x}_0}\\
        &= \delta \boldsymbol{x}_0^T\mathbf{\Phi }_{({t_0},t)}^T{\mathbf{\Phi }_{({t_0},t)}}\delta {\boldsymbol{x}_0} = \delta \boldsymbol{x}_0^T\boldsymbol{C}\delta {\boldsymbol{x}_0}
    \end{aligned} \,.
\end{equation}
Let ${\lambda _i}$ ($i \in \{ 1, \cdots ,n\} $) and ${\boldsymbol{\xi }_i} \in {\mathbb{R}^{n \times 1}}$ ($\left\| {{\boldsymbol{\xi }_i}} \right\| = 1$) be the \emph{i}-th eigenvalue and eigenvector pair of the CGT $\boldsymbol{C}$. These six eigenvalues are arranged in a descending order (\emph{i.e.}, ${\lambda _1} > {\lambda _2} >  \cdots  > {\lambda _{n - 1}} > {\lambda _n}$). If the initial state deviation $\delta \boldsymbol{x}_{0}$ is along the \emph{i}-th eigenvector ${\boldsymbol{\xi }_i}$ (\emph{i.e.}, $\delta {\boldsymbol{x}_0} = \left\| {\delta {\boldsymbol{x}_0}} \right\|{\boldsymbol{\xi }_i}$), Eq.~\eqref{eq:Norm of dx} can be simplified as
\begin{equation} \label{eq:Norm of dx (simplified)}
    {\left\| {\delta \boldsymbol{x}} \right\|^2} = {\left\| {\delta {\boldsymbol{x}_0}} \right\|^2}\boldsymbol{\xi }_i^T\boldsymbol{C}{\boldsymbol{\xi }_i} = {\left\| {\delta {\boldsymbol{x}_0}} \right\|^2}{\lambda _i} \,.
\end{equation}
One can see from Eq.~\eqref{eq:Norm of dx (simplified)} that, if the magnitude of the initial state deviation $\delta {\boldsymbol{x}_0}$ is fixed (\emph{i.e.}, $\left\| {\delta {\boldsymbol{x}_0}} \right\| = {\rm{constant}}$), an initial state deviation $\delta {\boldsymbol{x}_0}$ that is strictly along the eigenvector with the largest eigenvalue (\emph{i.e.}, $\delta {\boldsymbol{x}_0} = \left\| {\delta {\boldsymbol{x}_0}} \right\|{\boldsymbol{\xi }_1}$) will maximize the magnitude of the state deviation at epoch $t$ (\emph{i.e.}, maximize $\left\| {\delta \boldsymbol{x}} \right\|$). Thus, it is reasonable to use the eigenvectors of the CGT as sensitive directions. If only one eigenvector is used, the linear transformation matrix $\boldsymbol{R}$ can be written as
\begin{equation} \label{eq:R1}
    \boldsymbol{R} = \boldsymbol{\xi }_1^T \in {\mathbb{R}^{1 \times n}} \,,
\end{equation}
and if $m$ eigenvectors are employed, Eq.~\eqref{eq:R1} is rewritten as \cite{Boone2022JGCD}
\begin{equation} \label{eq:Rm}
    \boldsymbol{R} = [\boldsymbol{\xi }_1^T,\boldsymbol{\xi }_2^T, \cdots ,\boldsymbol{\xi }_m^T] \in {\mathbb{R}^{m \times n}} \,.
\end{equation}

The DSTT is efficient when applied to known orbits whose sensitive directions can be obtained in advance (\emph{e.g.}, a library of sensitive directions has been established \cite{Boodram2022, Boone2022, Boone2024} or the sensitive directions have been analytically revealed \cite{QIAO2024CJA}). For a new target orbit without prior information about its sensitive directions, there are usually two ways, called direct and indirect ways in this paper, to get the DSTTs. Using the direct way, one should first integrate the STM to directly compute the sensitive directions (using Eqs.~\eqref{eq:R1}-\eqref{eq:Rm}), and then integrate the DSTT (also contains the STM) using Eqs.~\eqref{eq:DSTT1 ODE}-\eqref{eq:DSTT4 ODE}. Otherwise, using the indirect way, one should integrate the full STT and then calculate the DSTT using its definition (\emph{i.e.}, Eqs.~\eqref{eq:DSTT}-\eqref{eq:DSTT4}). The direct way requires the repeated integration of the nominal orbit and the STM. The major drawback of the direct way lies in that one can only perform the orbital uncertainty propagation analysis at the predefined final epoch, as the linear transformation matrix $\boldsymbol{R}$ is constant when integrating the DSTTs (as shown in Eqs.~\eqref{eq:DSTT}-\eqref{eq:DSTT4}). 
To be specific, if the DSTTs are integrated from an initial epoch $t_0$ to a predefined final epoch $t_f$ and the terms $\varphi _{({t_0},{t_f})}^{i,{p_1} \cdots {p_p}}$ are obtained, one can only perform analysis at the epoch $t_f$. This is because the sensitive directions of the DSTT $\varphi _{({t_0},{t_f})}^{i,{p_1} \cdots {p_p}}$ is obtained based on the CGT at the predefined final epoch $t_f$; thus, these sensitive directions are valid at the epoch $t_f$ only.
If one wants to predict orbital uncertainties at a different epoch $t_{f}'$ ($t_{f}' \neq t_f$), a new DSTT (\emph{i.e.}, $\varphi _{({t_0},{t_f'})}^{i,{p_1} \cdots {p_p}}$) should be re-integrated from $t_0$ to $t_{f}'$ (based on the CGT at the epoch $t_{f}'$).
The indirect way can perform analysis at any time as the full STT doesn't have that problem; however, it has a slightly higher computational burden than the full STT as it includes the computation of the eigenvalue-eigenvector pairs of the CGT, which is contrary to the primary purpose of DSTT (\emph{i.e.}, to improve the computational efficiency of the full STT).

In the following section, a method called TDSTT is introduced, which can avoid the drawbacks of DSTT. The TDSTT computes the sensitive directions (\emph{i.e.}, the linear transformation matrix $\boldsymbol{R}$) with time so that one can use it to predict orbital uncertainties at any point (like the indirect way). In addition, it can still significantly reduce the complexity of the full STT while keeping almost the same level of accuracy.

\section{Derivatives of Eigenvalues and Eigenvectors}
\label{Sec:Derivatives of Eigenvalues and Eigenvectors}

\subsection{Derivatives of Eigenvalues} 
\label{Sec:Derivatives of Eigenvalues}
The CGT $\boldsymbol{C}$ is a $n$-dimensional symmetric matrix with six nonrepeated real eigenvalues. In addition, its eigenvectors are orthogonal with each other \cite{Boone2022JGCD}. According to Eq.~\eqref{eq:CGT}, the elements of the CGT $\boldsymbol{C}$ can be expressed as
\begin{equation} \label{eq:CGT elements}
    {C^{{k_1},{k_2}}} = \phi _{({t_0},t)}^{i,{k_1}}\phi _{({t_0},t)}^{i,{k_2}} \,,
\end{equation}
and the derivative of the CGT $\boldsymbol{C}$ can be derived as
\begin{equation} \label{eq:CGT derivatives}
    {\dot C^{{k_1},{k_2}}} = (\phi _{({t_0},t)}^{i,{k_1}}\phi _{({t_0},t)}^{i,{k_2}})' = \dot \phi _{({t_0},t)}^{i,{k_1}}\phi _{({t_0},t)}^{i,{k_2}} + \phi _{({t_0},t)}^{i,{k_1}}\dot \phi _{({t_0},t)}^{i,{k_2}} = {A^{i,\alpha }}\phi _{({t_0},t)}^{\alpha ,{k_1}}\phi _{({t_0},t)}^{i,{k_2}} + \phi _{({t_0},t)}^{i,{k_1}}{A^{i,\alpha }}\phi _{({t_0},t)}^{\alpha ,{k_2}} \,.
\end{equation}

Using the property of the eigenvalues and eigenvectors, one has
\begin{equation} \label{eq:eigenvalues and eigenvectors}
    \boldsymbol{C}{\boldsymbol{\xi }_k} = {\lambda _k}{\boldsymbol{\xi }_k} \,,
\end{equation}
which can be rewritten using the Einstein summation notation as
\begin{equation} \label{eq:eigenvalues and eigenvectors (notation)}
    {C^{i,j}}\xi _k^j = \sum\limits_{j = 1}^n {{C^{i,j}}\xi _k^j}  = {\lambda _k}\xi _k^i \,.
\end{equation}
where $\xi _k^i$ and $\xi _k^j$ denote the \emph{i}-th and \emph{j}-th elements of the \emph{k}-th eigenvector ${\boldsymbol{\xi }_k}$, respectively.

Derivating the equality $\boldsymbol{C}{\boldsymbol{\xi }_k} = {\lambda _k}{\boldsymbol{\xi }_k}$ (\emph{i.e.}, Eq.~\eqref{eq:eigenvalues and eigenvectors}) with respect to $t$, one can deduce the following equation for the \emph{k}-th eigenvector:
\begin{equation} \label{eq:eigenvalue1}
    \boldsymbol{\dot C}{\boldsymbol{\xi }_k} + \boldsymbol{C}{\boldsymbol{\dot \xi }_k} = {\dot \lambda _k}{\boldsymbol{\xi }_k} + {\lambda _k}{\boldsymbol{\dot \xi }_k} \,.
\end{equation}
Recall that the magnitude of the eigenvectors is equal to 1 (\emph{i.e.}, $\left\| {{\boldsymbol{\xi }_k}} \right\| = 1$); hence, one has
\begin{equation} \label{eq:eigenvalue2}
    {\boldsymbol{\dot \xi }_k} \bot {\boldsymbol{\xi }_k} \,,
\end{equation}
which indicates that the inner product of the eigenvector ${\boldsymbol{\xi }_k}$ and its derivative ${\boldsymbol{\dot \xi }_k}$ is zero, \emph{i.e.}, $({\boldsymbol{\dot \xi }_k},{\boldsymbol{\xi }_k} ) = \boldsymbol{\dot \xi }_k^T{\boldsymbol{\xi }_k} = 0$. Take the inner product of the Eq.~\eqref{eq:eigenvalue1} with the \emph{k}-th eigenvector, and one deduce
\begin{equation} \label{eq:eigenvalue3}
    (\boldsymbol{\dot C}{\boldsymbol{\xi }_k},{\boldsymbol{\xi }_k}) + (\boldsymbol{C}{\boldsymbol{\dot \xi }_k},{\boldsymbol{\xi }_k}) = ({\dot \lambda _k}{\boldsymbol{\xi }_k},{\boldsymbol{\xi }_k}) + ({\lambda _k}{\boldsymbol{\dot \xi }_k},{\boldsymbol{\xi }_k}) \,,
\end{equation}
Notice that $(\boldsymbol{C}{\boldsymbol{\dot \xi }_k},{\boldsymbol{\xi }_k}) = ({\boldsymbol{\dot \xi }_k},\boldsymbol{C}{\boldsymbol{\xi }_k}) = {\lambda _k}({\boldsymbol{\dot \xi }_k},{\boldsymbol{\xi }_k}) = 0$, and then, Eq.~\eqref{eq:eigenvalue3} can be further simplified as
\begin{equation} \label{eq:eigenvalue ODE}
    ({\dot \lambda _k}{\boldsymbol{\xi }_k},{\boldsymbol{\xi }_k}) = {\dot \lambda _k}{\left\| {{\boldsymbol{\xi }_k}} \right\|^2} = {\dot \lambda _k} = (\boldsymbol{\dot C}{\boldsymbol{\xi }_k},{\boldsymbol{\xi }_k}) \,,
\end{equation}
which determines the derivative of the \emph{k}-th eigenvalue.

\subsection{Derivatives of Eigenvectors} 
\label{Sec:Derivatives of Eigenvectors}
Two methods are introduced for computing the derivatives of eigenvectors. Both methods will be employed during the derivation of TDSTT.

\subsubsection{Linear representation method}
\label{Sec:First method for computing the eigenvector derivatives}
Taking the inner product of the Eq.~\eqref{eq:eigenvalue1} with the \emph{p}-th (satisfying $p \ne k$) eigenvector, one has
\begin{equation} \label{eq:eigenvector1}
    (\boldsymbol{\dot C}{\boldsymbol{\xi }_k},{\boldsymbol{\xi }_p}) + (\boldsymbol{C}{\boldsymbol{\dot \xi }_k},{\boldsymbol{\xi }_p}) = ({\dot \lambda _k}{\boldsymbol{\xi }_k},{\boldsymbol{\xi }_p}) + ({\lambda _k}{\boldsymbol{\dot \xi }_k},{\boldsymbol{\xi }_p}) \,.
\end{equation}
As eigenvectors are orthogonal, it follows that $({\dot \lambda _k}{\boldsymbol{\xi }_k},{\boldsymbol{\xi }_p}) = 0$, and using the property of symmetric matrices, Eq.~\eqref{eq:eigenvector1} becomes
\begin{equation} \label{eq:eigenvector2}
    (\boldsymbol{\dot C}{\boldsymbol{\xi }_k},{\boldsymbol{\xi }_p}) + ({\boldsymbol{\dot \xi }_k},\boldsymbol{C}{\boldsymbol{\xi }_p}) = ({\lambda _k}{\boldsymbol{\dot \xi }_k},{\boldsymbol{\xi }_p}) \,.
\end{equation}
Substituting the equation $\boldsymbol{C}{\boldsymbol{\xi }_p} = {\lambda _p}{\boldsymbol{\xi }_p}$ into Eq.~\eqref{eq:eigenvector2}, one has
\begin{equation} \label{eq:eigenvector3}
    (\boldsymbol{\dot C}{\boldsymbol{\xi }_k},{\boldsymbol{\xi }_p}) + {\lambda _p}({\boldsymbol{\dot \xi }_k},{\boldsymbol{\xi }_p}) = {\lambda _k}({\boldsymbol{\dot \xi }_k},{\boldsymbol{\xi }_p}) \,,
\end{equation}
which further yields
\begin{equation} \label{eq:eigenvector4}
    ({\boldsymbol{\dot \xi }_k},{\boldsymbol{\xi }_p}) = \frac{1}{{{\lambda _p} - {\lambda _k}}}(\boldsymbol{\dot C}{\boldsymbol{\xi }_k},{\boldsymbol{\xi }_p}) \,.
\end{equation}
Thus, one can determine the derivative of the \emph{k}-th eigenvector using the following equation:
\begin{equation} \label{eq:eigenvector5}
    {\boldsymbol{\dot \xi }_k} = \sum\limits_{p \ne k} {\frac{1}{{{\lambda _p} - {\lambda _k}}}(\boldsymbol{\dot C}{\boldsymbol{\xi }_k},{\boldsymbol{\xi }_p}){\boldsymbol{\xi }_p}} \,.
\end{equation}

Equation~\eqref{eq:eigenvector5} shows that the derivative of one eigenvector can be represented as a linear combination of the remaining eigenvectors, that is
\begin{equation} \label{eq:eigenvector B}
    {\boldsymbol{\dot \xi }_k} = {B^{k,p}}{\boldsymbol{\xi }_k} \,,
\end{equation}
where
\begin{equation} \label{eq:B}
    {B^{k,p}} = \left\{ {\begin{array}{*{20}{l}}
                    0 & {k = p}\\
                    {\frac{1}{{{\lambda _p} - {\lambda _k}}}(\boldsymbol{\dot C}{\boldsymbol{\xi }_k},{\boldsymbol{\xi }_p})} & {k \ne p}
                \end{array}} 
                \right. \,.
\end{equation}

One benefit of the above linear representation lies in that it can build a linear relationship between the eigenvector derivative and eigenvectors. However, its drawback is that it requires the knowledge of all eigenvalues and eigenvectors to generate the derivative of one eigenvector. In this work, this method is not used to generate eigenvector derivatives; its linear property is employed to simplify the expressions of TDSTT in the next section.

\subsubsection{Nelson’s method}
\label{Sec:Second method for computing the eigenvector derivatives}
A second method is introduced to calculate the eigenvector derivatives herein. The method was proposed by Nelson in 1976 \cite{Nelson1976}. For the \emph{k}-th eigenvalue and eigenvector pair $({\lambda _k},{\boldsymbol{\xi }_k})$, one should first define a vector $\boldsymbol{M}$ and a matrix $\boldsymbol{G}$ as
\begin{equation} \label{eq:M}
    \boldsymbol{M} = ({\dot \lambda _k}{\boldsymbol{I}_n} - \boldsymbol{\dot C}){\boldsymbol{\xi }_k} \,,
\end{equation}
\begin{equation} \label{eq:G}
    \boldsymbol{G} = \boldsymbol{C} - {\lambda _k}{\boldsymbol{I}_n} \,,
\end{equation}
where ${\boldsymbol{I}_n}$ represents an identified matrix of size $n \times n$. Then, find the index corresponding to the largest element of the \emph{k}-th eigenvector (\emph{i.e.}, find $i$ such that $\left| {\xi _k^i} \right| = {\left\| {{\boldsymbol{\xi }_k}} \right\|_\infty }$). Next, replacing the \emph{i}-th row and column of the matrix $\boldsymbol{G}$ with zero, replacing the \emph{i}-th diagonal element of the matrix $\boldsymbol{G}$ with 1, replacing the \emph{i}-th element of the vector $\boldsymbol{M}$ with 0, one has $\boldsymbol{\tilde G}$ and $\boldsymbol{\tilde M}$. Solve the following equation:
\begin{equation} \label{eq:v}
    \boldsymbol{\tilde Gv} = \boldsymbol{\tilde M} \,,
\end{equation}
and compute:
\begin{equation} \label{eq:c}
    c =  - (\boldsymbol{v},{\boldsymbol{\xi }_k}) \,.
\end{equation}
Finally, one can determine the derivative of the \emph{k}-th eigenvector as
\begin{equation} \label{eq:eigenvector ODE}
    {\boldsymbol{\dot \xi }_k} = \boldsymbol{v} + c{\boldsymbol{\xi }_k} \,,
\end{equation}

Nelson’s method has an attractive property of requiring only one eigenvector-eigenvalue pair (the one to be differentiated) to generate the eigenvector derivative \cite{Dailey1989}. This means that there is no need to calculate all the eigenvector-eigenvalue pairs when only several sensitive directions are focused on. In this work, Nelson’s method is directly employed to generate the eigenvector derivatives.

Note that both methods work only when the eigenvalues are nonrepeated and suffer from singularity when the matrix has repeated eigenvalues. For example, if one uses the first method (\emph{i.e.}, the linear representation method) to produce the eigenvector derivatives, repeated eigenvalues will lead to a zero denominator in Eqs.~\eqref{eq:eigenvector5}-\eqref{eq:B}. Some solutions have been proposed to handle the cases with repeated eigenvalues, including Ojalvo’s method \cite{Ojalvo1986} and Dailey’s method \cite{Dailey1989}.

\section{Time-varying Directional State Transition Tensor}
\label{Sec:Time-varying Directional State Transition Tensor}

\subsection{Basic Idea} \label{Sec:Basic Idea}
The STT technique requires the integration of the orbital state, first-order STT (\emph{i.e.}, STM), and higher-order STTs. The variable pair is given as $\{ \boldsymbol{x},\phi _{({t_0},t)}^{i,{k_1}},\phi _{({t_0},t)}^{i,{k_1}{k_2}}, \cdots \} $. The DSTT, once the linear transformation matrix $\boldsymbol{R}$ is determined in advance, requires the integration of the following variable pair: $\{ \boldsymbol{x},\phi _{({t_0},t)}^{i,{k_1}},\varphi _{({t_0},t)}^{i,{p_1}{p_2}}, \cdots \} $. To handle the case when the linear transformation matrix $\boldsymbol{R}$ is not available, this section introduces a time-varying DSTT technique, referred to as TDSTT. The basic idea of the TDSTT is to integrate the eigenvalues and eigenvectors of the CGT along the nominal orbit; therefore, the variable pair to be integrated is $\{ \boldsymbol{x},{\lambda _1}, \cdots ,{\lambda _m},{\boldsymbol{\xi }_1}, \cdots ,{\boldsymbol{\xi }_m},\phi _{({t_0},t)}^{i,{k_1}},\varphi _{({t_0},t)}^{i,{p_1}{p_2}}, \cdots \}$. The derivative of the orbital state is determined by Eq.~\eqref{eq:dynamics}, and the derivatives of the $m$ eigenvalue-eigenvector pairs are provided in Sec.~\ref{Sec:Derivatives of Eigenvalues and Eigenvectors}. One of the major differences between the DSTT and TDSTT lies in the linear transformation matrix $\boldsymbol{R}$. In the DSTT, the linear transformation matrix $\boldsymbol{R}$ is considered a constant matrix, whereas in this work (\emph{i.e.}, TDSTT), the linear transformation matrix $\boldsymbol{R}$ is time-dependent. In this case, Eqs.~\eqref{eq:DSTT1 ODE}-\eqref{eq:DSTT4 ODE} no longer hold, and new equations are derived in the following subsection.

\subsection{Derivatives of TDSTT} \label{Sec:Derivatives of TDSTT}
According to Eqs.~\eqref{eq:DSTT}-\eqref{eq:DSTT4}, the TDSTTs rely on both the full STTs and the linear transformation matrix $\boldsymbol{R}$. However, during the integration of the TDSTTs, the full STTs (except for the first-order STT) are not available. To overcome this problem, a method is first provided to approximate the full STT using TDSTTs herein. The expressions for approximating the first- and second-order full STTs are given as
\begin{equation} \label{eq:DSTT-STT1}
    \phi _{({t_0},t)}^{i,{k_1}} = \frac{{\partial {x^i}}}{{\partial x_0^{{k_1}}}} \approx \frac{{\partial {x^i}}}{{\partial y_0^{{p_1}}}}\frac{{\partial y_0^{{p_1}}}}{{\partial x_0^{{k_1}}}} = \varphi _{({t_0},t)}^{i,{p_1}}{R^{{p_1},{k_1}}} \,,
\end{equation}
\begin{equation} \label{eq:DSTT-STT2}
    \phi _{({t_0},t)}^{i,{k_1}{k_2}} = \frac{{{\partial ^2}{x^i}}}{{\partial x_0^{{k_1}}\partial x_0^{{k_2}}}} \approx \frac{{{\partial ^2}{x^i}}}{{\partial y_0^{{p_1}}\partial y_0^{{p_2}}}}\frac{{\partial y_0^{{p_1}}}}{{\partial x_0^{{k_1}}}}\frac{{\partial y_0^{{p_2}}}}{{\partial x_0^{{k_2}}}} = \varphi _{({t_0},t)}^{i,{p_1}{p_2}}{R^{{p_1},{k_1}}}{R^{{p_2},{k_2}}} \,,
\end{equation}
and the approximated expression of a \emph{p}-th order full STT is derived as
\begin{equation} \label{eq:DSTT-STT}
    \phi _{({t_0},t)}^{i,{k_1} \cdots {k_p}} = \frac{{{\partial ^p}{x^i}}}{{\partial x_0^{{k_1}} \cdots \partial x_0^{{k_p}}}} \approx \frac{{{\partial ^p}{x^i}}}{{\partial y_0^{{p_1}} \cdots \partial y_0^{{p_p}}}}\frac{{\partial y_0^{{p_1}}}}{{\partial x_0^{{k_1}}}} \cdots \frac{{\partial y_0^{{p_p}}}}{{\partial x_0^{{k_p}}}} = \varphi _{({t_0},t)}^{i,{p_1} \cdots {p_p}}{R^{{p_1},{k_1}}} \cdots {R^{{p_p},{k_p}}} \,.
\end{equation}

As the first-order STT (STM) is preserved, this work focuses on deriving the derivatives of the second- and higher-order TDSTTs. First, the derivative of the second-order TDSTT $\varphi _{({t_0},t)}^{i,{p_1}{p_2}}$ will be explicitly written out. According to Eq.~\eqref{eq:DSTT}, the differential equation of $\varphi _{({t_0},t)}^{i,{p_1}{p_2}}$ can be written as
\begin{equation} \label{eq:TDSTT2 ODE1}
    \begin{aligned}
        \dot \varphi _{({t_0},t)}^{i,{p_1}{p_2}} &= (\phi _{({t_0},t)}^{i,{k_1}{k_2}}{R^{{p_1},{k_1}}}{R^{{p_2},{k_2}}})'\\
        &= \dot \phi _{({t_0},t)}^{i,{k_1}{k_2}}{R^{{p_1},{k_1}}}{R^{{p_2},{k_2}}} + \phi _{({t_0},t)}^{i,{k_1}{k_2}}{{\dot R}^{{p_1},{k_1}}}{R^{{p_2},{k_2}}} + \phi _{({t_0},t)}^{i,{k_1}{k_2}}{R^{{p_1},{k_1}}}{{\dot R}^{{p_2},{k_2}}}\\
        &= {A^{i,\alpha }}\varphi _{({t_0},t)}^{\alpha ,{p_1}{p_2}} + {A^{i,\alpha \beta }}\varphi _{({t_0},t)}^{\alpha ,{p_1}}\varphi _{({t_0},t)}^{\beta ,{p_2}} + \phi _{({t_0},t)}^{i,{k_1}{k_2}}{{\dot R}^{{p_1},{k_1}}}{R^{{p_2},{k_2}}} + \phi _{({t_0},t)}^{i,{k_1}{k_2}}{R^{{p_1},{k_1}}}{{\dot R}^{{p_2},{k_2}}}
    \end{aligned} \,.
\end{equation}
Substituting Eq.~\eqref{eq:DSTT-STT2} into Eq.~\eqref{eq:TDSTT2 ODE1}, one can write
\begin{equation} \label{eq:TDSTT2 ODE2}
    \dot \varphi _{({t_0},t)}^{i,{p_1}{p_2}} = {A^{i,\alpha }}\varphi _{({t_0},t)}^{\alpha ,{p_1}{p_2}} + {A^{i,\alpha \beta }}\varphi _{({t_0},t)}^{\alpha ,{p_1}}\varphi _{({t_0},t)}^{\beta ,{p_2}} + \varphi _{({t_0},t)}^{i,{\gamma _1}{\gamma _2}}{R^{{\gamma _1},{k_1}}}{R^{{\gamma _2},{k_2}}}{\dot R^{{p_1},{k_1}}}{R^{{p_2},{k_2}}} + \varphi _{({t_0},t)}^{i,{\gamma _1}{\gamma _2}}{R^{{\gamma _1},{k_1}}}{R^{{\gamma _2},{k_2}}}{R^{{p_1},{k_1}}}{\dot R^{{p_2},{k_2}}} \,.
\end{equation}

Note that the eigenvectors are orthogonal; thus, one has
\begin{equation} \label{eq:TDSTT2 ODE3}
    {R^{{\gamma _2},{k_2}}}{R^{{p_2},{k_2}}} = \left\{ {\begin{array}{*{20}{l}}
        1&{{\gamma _2} = {p_2}}\\
        0&{{\gamma _2} \ne {p_2}}
    \end{array}} \right. \,.
\end{equation}
Using Eqs.~\eqref{eq:DSTT-STT2} and \eqref{eq:TDSTT2 ODE3}, the third term on the right side of Eq.~\eqref{eq:TDSTT2 ODE1} can be written as
\begin{equation} \label{eq:TDSTT2 ODE4}
    \phi _{({t_0},t)}^{i,{k_1}{k_2}}{\dot R^{{p_1},{k_1}}}{R^{{p_2},{k_2}}} = \varphi _{({t_0},t)}^{i,{\gamma _1}{\gamma _2}}{R^{{\gamma _1},{k_1}}}{R^{{\gamma _2},{k_2}}}{\dot R^{{p_1},{k_1}}}{R^{{p_2},{k_2}}} = \varphi _{({t_0},t)}^{i,{\gamma _1}{p_2}}{R^{{\gamma _1},{k_1}}}{\dot R^{{p_1},{k_1}}} \,.
\end{equation}
Then, using Eq.~\eqref{eq:eigenvector B} (\emph{i.e.}, the linear representation method for computing the eigenvector derivatives), one has
\begin{equation} \label{eq:TDSTT2 ODE5}
    {\dot R^{{p_1},{k_1}}} = {B^{{p_1},p}}{R^{p,{k_1}}} \,.
\end{equation}
Substitution of Eq.~\eqref{eq:TDSTT2 ODE5} into Eq.~\eqref{eq:TDSTT2 ODE4} yields
\begin{equation} \label{eq:TDSTT2 ODE6}
    \varphi _{({t_0},t)}^{i,{\gamma _1}{p_2}}{R^{{\gamma _1},{k_1}}}{\dot R^{{p_1},{k_1}}} = \varphi _{({t_0},t)}^{i,{\gamma _1}{p_2}}{R^{{\gamma _1},{k_1}}}{B^{{p_1},p}}{R^{p,{k_1}}} = \varphi _{({t_0},t)}^{i,{\gamma _1}{p_2}}{B^{{p_1},{\gamma _1}}} \,.
\end{equation}

Similarly, the fourth term on the right side of Eq.~\eqref{eq:TDSTT2 ODE1} can be simplified as
\begin{equation} \label{eq:TDSTT2 ODE7}
    \begin{aligned}
        \phi _{({t_0},t)}^{i,{k_1}{k_2}}{R^{{p_1},{k_1}}}{{\dot R}^{{p_2},{k_2}}} &= \varphi _{({t_0},t)}^{i,{\gamma _1}{\gamma _2}}{R^{{\gamma _1},{k_1}}}{R^{{\gamma _2},{k_2}}}{R^{{p_1},{k_1}}}{{\dot R}^{{p_2},{k_2}}}\\
        &= \varphi _{({t_0},t)}^{i,{p_1}{\gamma _2}}{R^{{\gamma _2},{k_2}}}{{\dot R}^{{p_2},{k_2}}}\\
        &= \varphi _{({t_0},t)}^{i,{p_1}{\gamma _2}}{B^{{p_2},{\gamma _2}}}
    \end{aligned} \,.
\end{equation}
Substituting Eqs.~\eqref{eq:TDSTT2 ODE6}-\eqref{eq:TDSTT2 ODE7} into Eq.~\eqref{eq:TDSTT2 ODE1}, one finally has
\begin{equation} \label{eq:TDSTT2 ODE}
    \dot \varphi _{({t_0},t)}^{i,{p_1}{p_2}} = {A^{i,\alpha }}\varphi _{({t_0},t)}^{\alpha ,{p_1}{p_2}} + {A^{i,\alpha \beta }}\varphi _{({t_0},t)}^{\alpha ,{p_1}}\varphi _{({t_0},t)}^{\beta ,{p_2}} + \varphi _{({t_0},t)}^{i,{\gamma _1}{p_2}}{B^{{p_1},{\gamma _1}}} + \varphi _{({t_0},t)}^{i,{p_1}{\gamma _2}}{B^{{p_2},{\gamma _2}}} \,,
\end{equation}
which determines the derivatives of the second-order TDSTT. Using Eq.~\eqref{eq:TDSTT2 ODE}, the complexity of computing the derivative of a second-order TDSTT has reduced from $\mathrm{O}(n+n^2+2n^4m^2)$ to $\mathrm{O}(n+n^2+2m)$. One can see from Eq.~\eqref{eq:TDSTT2 ODE} that the first two terms of the TDSTT derivative (\emph{i.e.}, ${A^{i,\alpha }}\varphi _{({t_0},t)}^{\alpha ,{p_1}{p_2}} + {A^{i,\alpha \beta }}\varphi _{({t_0},t)}^{\alpha ,{p_1}}\varphi _{({t_0},t)}^{\beta ,{p_2}}$) are exactly the same as the derivative of the second-order DSTT (\emph{i.e.}, Eq.~\eqref{eq:DSTT2 ODE}). The changes in the time-dependent linear transformation matrix $\boldsymbol{R}$ lead to the remaining terms in Eq.~\eqref{eq:TDSTT2 ODE} (\emph{i.e.}, $\varphi _{({t_0},t)}^{i,{\gamma _1}{p_2}}{B^{{p_1},{\gamma _1}}}$ and $\varphi _{({t_0},t)}^{i,{p_1}{\gamma _2}}{B^{{p_2},{\gamma _2}}}$).

Then, the derivative of the third-order TDSTT is derived as follows. Using the definition of the third-order TDSTT (\emph{i.e.}, $\varphi _{({t_0},t)}^{i,{p_1}{p_2}{p_3}} = \phi _{({t_0},t)}^{i,{k_1}{k_2}{k_3}}{R^{{p_1},{k_1}}}{R^{{p_2},{k_2}}}{R^{{p_3},{k_3}}}$), one can deduce its derivative as
\begin{equation} \label{eq:TDSTT3 ODE1}
    \begin{aligned}
        \dot \varphi _{({t_0},t)}^{i,{p_1}{p_2}{p_3}} &= (\phi _{({t_0},t)}^{i,{k_1}{k_2}{k_3}}{R^{{p_1},{k_1}}}{R^{{p_2},{k_2}}}{R^{{p_3},{k_3}}})'\\
        &= \dot \phi _{({t_0},t)}^{i,{k_1}{k_2}{k_3}}{R^{{p_1},{k_1}}}{R^{{p_2},{k_2}}}{R^{{p_3},{k_3}}} + \phi _{({t_0},t)}^{i,{k_1}{k_2}{k_3}}{{\dot R}^{{p_1},{k_1}}}{R^{{p_2},{k_2}}}{R^{{p_3},{k_3}}}\\
        &+ \phi _{({t_0},t)}^{i,{k_1}{k_2}{k_3}}{R^{{p_1},{k_1}}}{{\dot R}^{{p_2},{k_2}}}{R^{{p_3},{k_3}}} + \phi _{({t_0},t)}^{i,{k_1}{k_2}{k_3}}{R^{{p_1},{k_1}}}{R^{{p_2},{k_2}}}{{\dot R}^{{p_3},{k_3}}}
    \end{aligned} \,.
\end{equation}
Using Eqs.~\eqref{eq:STT3 ODE}, \eqref{eq:TDSTT2 ODE3}, and \eqref{eq:TDSTT2 ODE5}, one has
\begin{equation} \label{eq:TDSTT3 ODE2}
    \begin{aligned}
        \dot \phi _{({t_0},t)}^{i,{k_1}{k_2}{k_3}}{R^{{p_1},{k_1}}}{R^{{p_2},{k_2}}}{R^{{p_3},{k_3}}} &= {A^{i,\alpha \beta }}\left[ {\varphi _{({t_0},t)}^{\alpha ,{p_1}}\varphi _{({t_0},t)}^{\beta ,{p_2}{p_3}} + \varphi _{({t_0},t)}^{\alpha ,{p_1}{p_2}}\varphi _{({t_0},t)}^{\beta ,{p_3}} + \varphi _{({t_0},t)}^{\alpha ,{p_1}{p_3}}\varphi _{({t_0},t)}^{\beta ,{p_2}}} \right]\\
        &+ {A^{i,\alpha }}\varphi _{({t_0},t)}^{\alpha ,{p_1}{p_2}{p_3}} + {A^{i,\alpha \beta \lambda }}\varphi _{({t_0},t)}^{\alpha ,{p_1}}\varphi _{({t_0},t)}^{\beta ,{p_2}}\varphi _{({t_0},t)}^{\lambda ,{p_3}}
    \end{aligned} \,,
\end{equation}
\begin{equation} \label{eq:TDSTT3 ODE3}
    \begin{aligned}
        \phi _{({t_0},t)}^{i,{k_1}{k_2}{k_3}}{{\dot R}^{{p_1},{k_1}}}{R^{{p_2},{k_2}}}{R^{{p_3},{k_3}}} &= \varphi _{({t_0},t)}^{i,{\gamma _1}{\gamma _2}{\gamma _3}}{R^{{\gamma _1},{k_1}}}{R^{{\gamma _2},{k_2}}}{R^{{\gamma _3},{k_3}}}{{\dot R}^{{p_1},{k_1}}}{R^{{p_2},{k_2}}}{R^{{p_3},{k_3}}}\\
        &= \varphi _{({t_0},t)}^{i,{\gamma _1}{p_2}{p_3}}{R^{{\gamma _1},{k_1}}}{{\dot R}^{{p_1},{k_1}}}\\
        &= \varphi _{({t_0},t)}^{i,{\gamma _1}{p_2}{p_3}}{R^{{\gamma _1},{k_1}}}{B^{{p_1},p}}{R^{p,{k_1}}}\\
        &= \varphi _{({t_0},t)}^{i,{\gamma _1}{p_2}{p_3}}{B^{{p_1},{\gamma _1}}}
    \end{aligned} \,,
\end{equation}
\begin{equation} \label{eq:TDSTT3 ODE4}
    \phi _{({t_0},t)}^{i,{k_1}{k_2}{k_3}}{R^{{p_1},{k_1}}}{\dot R^{{p_2},{k_2}}}{R^{{p_3},{k_3}}} = \varphi _{({t_0},t)}^{i,{p_1}{\gamma _2}{p_3}}{B^{{p_2},{\gamma _2}}} \,,
\end{equation}
\begin{equation} \label{eq:TDSTT3 ODE5}
    \phi _{({t_0},t)}^{i,{k_1}{k_2}{k_3}}{R^{{p_1},{k_1}}}{R^{{p_2},{k_2}}}{\dot R^{{p_3},{k_3}}} = \varphi _{({t_0},t)}^{i,{p_1}{p_2}{\gamma _3}}{B^{{p_3},{\gamma _3}}} \,.
\end{equation}

Substitution of Eqs.~\eqref{eq:TDSTT3 ODE2}-\eqref{eq:TDSTT3 ODE5} into Eq.~\eqref{eq:TDSTT3 ODE1} produces a simplified version of the differential equation of the third-order TDSTT, given as
\begin{equation} \label{eq:TDSTT3 ODE}
    \begin{aligned} 
        \dot \varphi _{({t_0},t)}^{i,{p_1}{p_2}{p_3}} &= {A^{i,\alpha \beta }}\left[ {\varphi _{({t_0},t)}^{\alpha ,{p_1}}\varphi _{({t_0},t)}^{\beta ,{p_2}{p_3}} + \varphi _{({t_0},t)}^{\alpha ,{p_1}{p_2}}\varphi _{({t_0},t)}^{\beta ,{p_3}} + \varphi _{({t_0},t)}^{\alpha ,{p_1}{p_3}}\varphi _{({t_0},t)}^{\beta ,{p_2}}} \right]\\
        &+ {A^{i,\alpha }}\varphi _{({t_0},t)}^{\alpha ,{p_1}{p_2}{p_3}} + {A^{i,\alpha \beta \lambda }}\varphi _{({t_0},t)}^{\alpha ,{p_1}}\varphi _{({t_0},t)}^{\beta ,{p_2}}\varphi _{({t_0},t)}^{\lambda ,{p_3}}\\
        &+ \varphi _{({t_0},t)}^{i,{\gamma _1}{p_2}{p_3}}{B^{{p_1},{\gamma _1}}} + \varphi _{({t_0},t)}^{i,{p_1}{\gamma _2}{p_3}}{B^{{p_2},{\gamma _2}}} + \varphi _{({t_0},t)}^{i,{p_1}{p_2}{\gamma _3}}{B^{{p_3},{\gamma _3}}}
    \end{aligned} \,.
\end{equation}

Following the above procedures, the derivative of the fourth-order TDSTT is derived as
\begin{equation} \label{eq:TDSTT4 ODE}
    \begin{aligned}
        \dot \varphi _{({t_0},t)}^{i,{p_1}{p_2}{p_3}{p_4}} &= {A^{i,\alpha }}\varphi _{({t_0},t)}^{\alpha ,{p_1}{p_2}{p_3}{\gamma _4}} + {A^{i,\alpha \beta }}\left[ {\varphi _{({t_0},t)}^{\alpha ,{p_1}{p_2}{p_3}}\varphi _{({t_0},t)}^{\beta ,{p_4}} + \varphi _{({t_0},t)}^{\alpha ,{p_1}{p_2}{p_4}}\varphi _{({t_0},t)}^{\beta ,{p_3}} + \varphi _{({t_0},t)}^{\alpha ,{p_1}{p_3}{p_4}}\varphi _{({t_0},t)}^{\beta ,{p_2}}} \right.\\
        &+ \left. {\varphi _{({t_0},t)}^{\alpha ,{p_1}{p_2}}\varphi _{({t_0},t)}^{\beta ,{p_3}{p_4}} + \varphi _{({t_0},t)}^{\alpha ,{p_1}{p_3}}\varphi _{({t_0},t)}^{\beta ,{p_2}{p_4}} + \varphi _{({t_0},t)}^{\alpha ,{p_1}{p_4}}\varphi _{({t_0},t)}^{\beta ,{p_2}{p_3}} + \varphi _{({t_0},t)}^{\alpha ,{p_1}}\varphi _{({t_0},t)}^{\beta ,{p_2}{p_3}{p_4}}} \right]\\
        &+ {A^{i,\alpha \beta \lambda }}\left[ {\varphi _{({t_0},t)}^{\alpha ,{p_1}{p_2}}\varphi _{({t_0},t)}^{\beta ,{p_3}}\varphi _{({t_0},t)}^{\lambda ,{p_4}} + \varphi _{({t_0},t)}^{\alpha ,{p_1}{p_3}}\varphi _{({t_0},t)}^{\beta ,{p_2}}\varphi _{({t_0},t)}^{\lambda ,{p_4}} + \varphi _{({t_0},t)}^{\alpha ,{p_1}{p_4}}\varphi _{({t_0},t)}^{\beta ,{p_2}}\varphi _{({t_0},t)}^{\lambda ,{p_3}}} \right.\\
        &+ \left. {\varphi _{({t_0},t)}^{\alpha ,{p_1}}\varphi _{({t_0},t)}^{\beta ,{p_2}{p_3}}\varphi _{({t_0},t)}^{\lambda ,{p_4}} + \varphi _{({t_0},t)}^{\alpha ,{p_1}}\varphi _{({t_0},t)}^{\beta ,{p_2}{p_4}}\varphi _{({t_0},t)}^{\lambda ,{p_3}} + \varphi _{({t_0},t)}^{\alpha ,{p_1}}\varphi _{({t_0},t)}^{\beta ,{p_2}}\varphi _{({t_0},t)}^{\lambda ,{p_3}{p_4}}} \right]\\
        &+ {A^{i,\alpha \beta \lambda \delta }}\varphi _{({t_0},t)}^{\alpha ,{p_1}}\varphi _{({t_0},t)}^{\beta ,{p_2}}\varphi _{({t_0},t)}^{\lambda ,{p_3}}\varphi _{({t_0},t)}^{\delta ,{p_4}}\\
        &+ \varphi _{({t_0},t)}^{i,{\gamma _1}{p_2}{p_3}{p_4}}{B^{{p_1},{\gamma _1}}} + \varphi _{({t_0},t)}^{i,{p_1}{\gamma _2}{p_3}{p_4}}{B^{{p_2},{\gamma _2}}}\\
        &+ \varphi _{({t_0},t)}^{i,{p_1}{p_2}{\gamma _3}{p_4}}{B^{{p_3},{\gamma _3}}} + \varphi _{({t_0},t)}^{i,{p_1}{p_2}{p_3}{\gamma _4}}{B^{{p_4},{\gamma _4}}}
    \end{aligned} \,.
\end{equation}
The differential equations of higher-order TDSTTs can be easily derived using a similar procedure, which are not presented in this paper for the sake of simplicity.

\subsection{Overall Procedure} \label{Sec:Overall Procedure}
According to the discussions in Sec.~\ref{Sec:Basic Idea} and Sec.~\ref{Sec:Derivatives of TDSTT}, the overall procedure of the proposed TDSTT method is summarized herein \footnote{Note that the code (Python version) to implement this is available at \url{https://github.com/ZHOUXINGYU-OASIS/TDSTT}.}. To integrate the TDSTT, one must first calculate its derivatives. Different from the previous STTs and DSTTs, whose derivatives rely on only the local Jacobi element (see Eqs.~\eqref{eq:STT1 ODE}-\eqref{eq:STT4 ODE} and Eqs.~\eqref{eq:DSTT1 ODE}-\eqref{eq:DSTT4 ODE}, the differential equations of the proposed TDSTT require the knowledge of the derivatives of the CGT and its eigenvalue-eigenvector pairs. A pseudocode to compute the derivative of a second-order TDSTT is shown in Table~\ref{tab:Pseudocode to compute the derivative of a second-order TDSTT}. Recall that the two methods for computing the eigenvector derivatives are employed. As shown in Table~\ref{tab:Pseudocode to compute the derivative of a second-order TDSTT}, the first method (\emph{i.e.}, Eqs.~\eqref{eq:M}-\eqref{eq:eigenvector ODE} is engaged in \textbf{step 6}, whereas the second method (\emph{i.e.}, Eqs.~\eqref{eq:eigenvector B}-\eqref{eq:B} is partly used in \textbf{step 7}.

\begin{table}[!h]
    \caption{\label{tab:Pseudocode to compute the derivative of a second-order TDSTT} Pseudocode to compute the derivative of a second-order TDSTT}
    \centering
    \begin{tabular}{lr}
        \hline\hline
        \multicolumn{2}{l}{\textbf{procedure} [$\boldsymbol{\dot x},\dot \phi _{({t_0},t)}^{i,{k_1}},{\dot \lambda _k},{\boldsymbol{\dot \xi }_k},\dot \varphi _{({t_0},t)}^{i,{p_1}{p_2}}$] = \textbf{diffTDSTT}($\boldsymbol{x},\phi _{({t_0},t)}^{i,{k_1}},{\lambda _k},{\boldsymbol{\xi }_k},\varphi _{({t_0},t)}^{i,{p_1}{p_2}},m$)} \\
        \,\,\,\,\textbf{1}: Compute the derivative of the orbital state $\boldsymbol{x}$ & $\rhd$ Eq.~\eqref{eq:dynamics} \\
        \,\,\,\,\textbf{2}: Compute the derivative of the first-order STT $\phi _{({t_0},t)}^{i,{k_1}}$ & $\rhd$ Eq.~\eqref{eq:STT1 ODE} \\
        \,\,\,\,\textbf{3}: Compute the CGT $\boldsymbol{C}$ & $\rhd$ Eq.~\eqref{eq:CGT} \\
        \,\,\,\,\textbf{4}: Compute the derivative of the CGT $\boldsymbol{C}$ & $\rhd$ Eq.~\eqref{eq:CGT derivatives} \\
        \,\,\,\,\textbf{5}: Compute the derivatives of $m$ eigenvalues $\{{\lambda _1}, \cdots ,{\lambda _m}\}$ & $\rhd$ Eq.~\eqref{eq:eigenvalue ODE} \\
        \,\,\,\,\textbf{6}: Compute the derivatives of $m$ eigenvectors $\{{\boldsymbol{\xi }_1}, \cdots ,{\boldsymbol{\xi }_m}\}$ & $\rhd$ Eqs.~\eqref{eq:M}-\eqref{eq:eigenvector ODE} \\
        \,\,\,\,\textbf{7}: Compute the coefficients ${B^{k,p}}$ & $\rhd$ Eqs.~\eqref{eq:eigenvector B}-\eqref{eq:B} \\
        \,\,\,\,\textbf{8}: Compute the derivative of the second-order TDSTT $\varphi _{({t_0},t)}^{i,{p_1}{p_2}}$ & $\rhd$ Eq.~\eqref{eq:TDSTT2 ODE} \\
        \multicolumn{2}{l}{\textbf{return} $\boldsymbol{\dot x},\dot \phi _{({t_0},t)}^{i,{k_1}},{\dot \lambda _k},{\boldsymbol{\dot \xi }_k},\dot \varphi _{({t_0},t)}^{i,{p_1}{p_2}}$} \\
        \hline\hline
    \end{tabular}
\end{table}

One strategy employed in this work is to take the logarithm when integrating the eigenvalues to improve the efficiency of numerical integrations. The other thing that should be discussed is the singularity of the TDSTT at the initial epoch $t_0$. Note that the first-order STT (\emph{i.e.}, STM) is initialized as an identified matrix, leading to the CGT being an identified matrix at the initial epoch. It is well known that a $n$-dimensional identified matrix has $n$ repeated eigenvalues (\emph{i.e.}, all eigenvalues are equal to 1), and any vector can be the eigenvector of an identified matrix. Therefore, at the initial epoch, the eigenvalue-eigenvector derivatives cannot be correctly obtained using Eqs.~\eqref{eq:eigenvalue ODE} and \eqref{eq:eigenvector ODE}. A warm start strategy is proposed to handle such a problem. To be specific, if the orbital uncertainties are to be propagated from an initial epoch $t_0$ to a given future epoch $t$, one can first integrate the full STT for a small step (\emph{i.e.}, from $t_0$ to $t'$). Then, the CGT, as well as its eigenvalue-eigenvector pairs, can be easily computed at the epoch $t'$. Finally, the TDSTT can be integrated along the nominal orbit from the epoch $t'$ to the final epoch $t$. The pseudocode of computing a TDSTT is provided in Table~\ref{tab:Pseudocode to compute a TDSTT}. Once the TDSTT is integrated, the orbital state deviation $\delta \boldsymbol{x}$ can be predicted using Eq.~\eqref{eq:Taylor DSTT}. Additionally, the statical moments (such as the mean and covariance) of the orbital state can be obtained using TDSTT. The equations for obtaining the statical moments using TDSTT are exactly the same as the DSTT, which is not shown in this paper for the sake of simplicity. Readers with a particular interest can refer to Ref.~\cite{Boone2022JGCD}.

\begin{table}[!h]
    \caption{\label{tab:Pseudocode to compute a TDSTT} Pseudocode to compute a TDSTT}
    \centering
    \begin{tabular}{lr}
        \hline\hline
        \multicolumn{2}{l}{\textbf{procedure} [$\phi _{({t_0},t)}^{i,{k_1}},{\lambda _k},{\boldsymbol{\xi }_k},\varphi _{({t_0},t)}^{i,{p_1} \cdots {p_p}}$] = \textbf{TDSTT}(${\boldsymbol{x}_0},{t_0},t,t',m,P$)} \\
        \,\,\,\,\textbf{1}: Integrate the (full) STTs up to the $P$-th order from the epoch $t_0$ to $t'$ & $\rhd$ Eqs.~\eqref{eq:STT1 ODE}-\eqref{eq:STT4 ODE} \\
        \,\,\,\,\textbf{2}: Compute the CGT at epoch $t'$ & $\rhd$ Eq.~\eqref{eq:CGT} \\
        \,\,\,\,\textbf{3}: Compute the $m$ eigenvalue-eigenvector pairs of the CGT &  \\
        \,\,\,\,\textbf{4}: Construct the linear transformation matrix $\boldsymbol{R}$ & $\rhd$ Eqs.~\eqref{eq:R1}-\eqref{eq:Rm} \\
        \,\,\,\,\textbf{5}: Integrate the TDSTTs up to the $P$-th order from the epoch $t'$ to $t$ & $\rhd$ Eqs.~\eqref{eq:TDSTT2 ODE}, \eqref{eq:TDSTT3 ODE}, \eqref{eq:TDSTT4 ODE} \\
        \,\,\,\,\textbf{6}: Obtain $\phi _{({t_0},t)}^{i,{k_1}},{\lambda _k},{\boldsymbol{\xi }_k},\varphi _{({t_0},t)}^{i,{p_1} \cdots {p_p}}$ &  \\
        \multicolumn{2}{l}{\textbf{return} $\phi _{({t_0},t)}^{i,{k_1}},{\lambda _k},{\boldsymbol{\xi }_k},\varphi _{({t_0},t)}^{i,{p_1} \cdots {p_p}}$} \\
        \hline\hline
    \end{tabular}
\end{table}

\subsection{Complexity analysis} \label{Sec:Complexity analysis}
Table~\ref{tab:Variables to be integrated for different methods} shows the variables to be integrated using the (full) STT, DSTT, and TDSTT. For the DSTT, it is assumed that the sensitive directions are not available before integrating the STM, and the direct way (\emph{i.e.}, integrate the STM first to compute sensitive directions and then integrate the DSTT) is employed to obtain its terms. The final row of Table~\ref{tab:Variables to be integrated for different methods} presents the total number of variables required to be integrated, particularly the number of variables when $n=6$ (\emph{i.e.}, for a 3-dimensional trajectory) and $m=1$ (\emph{i.e.}, using only one sensitive direction) are listed inside the parenthesis. One can see from Table~\ref{tab:Variables to be integrated for different methods} that the second-order TDSTT only requires the numerical integration of $n+n^2+(n+1)m+nm^2$ variables, while the second-order full STT and DSTT require $n+n^2+n^3$ and $2n+2n^2+nm^2$ variables, respectively. To be specific, when $n=6$ and $m=1$, the second-order TDSTT (55 variables) requires 78.50\% fewer variables than the full STT (256 variables). The number of variables to be integrated grows exponentially ($n + \sum\limits_{p = 1}^P {{n^{p + 1}}}$) as the order of the STT increases, while the variable amount increases polynomially ($n + {n^2} + (n + 1)m + (P - 1)n$ if $m=1$) using the proposed TDSTT. Moreover, the TDSTT also requires fewer variables than the DSTT. It can save more than 38.88\% of variables at the order of two, and that percentage becomes 36.45\% when the order is three. The direct way of computing the DSTT can reduce the algorithm complexity of full STT, however, at the cost of only working at a predefined epoch.

\begin{table}[!h]
    \caption{\label{tab:Variables to be integrated for different methods} Variables to be integrated for different methods}
    \centering
    \begin{tabular}{cccccccc}
        \hline\hline
        \multirow{2}{*}{Variable} & $1^{\rm{st}}$ order & \multicolumn{3}{c}{$2^{\rm{nd}}$ order} & \multicolumn{3}{c}{$3^{\rm{rd}}$ order} \\ \cline{2-8}
        & STM & STT & DSTT & TDSTT & STT & DSTT & TDSTT \\ \hline
        $\boldsymbol{x}$ & ${\mathbb{R}^n}$ & ${\mathbb{R}^n}$ & ${\mathbb{R}^n} \times 2$ & ${\mathbb{R}^n}$ & ${\mathbb{R}^n}$ & ${\mathbb{R}^n} \times 2$ & ${\mathbb{R}^n}$ \\
        $\phi _{({t_0},t)}^{i,{k_1}}$ & ${\mathbb{R}^{n \times n}}$ & ${\mathbb{R}^{n \times n}}$ & ${\mathbb{R}^{n \times n}} \times 2$ & ${\mathbb{R}^{n \times n}}$ & ${\mathbb{R}^{n \times n}}$ & ${\mathbb{R}^{n \times n}} \times 2$ & ${\mathbb{R}^{n \times n}}$ \\
        ${\lambda _k}$ & $\times$ & $\times$ & $\times$ & ${\mathbb{R}^m}$ & $\times$ & $\times$ & ${\mathbb{R}^m}$ \\
        ${\boldsymbol{\xi }_k}$ & $\times$ & $\times$ & $\times$ & ${\mathbb{R}^{m \times n}}$ & $\times$ & $\times$ & ${\mathbb{R}^{m \times n}}$ \\
        $\phi _{({t_0},t)}^{i,{k_1}{k_2}}$ & $\times$ & ${\mathbb{R}^{n \times n \times n}}$ & ${\mathbb{R}^{n \times n \times n}}$ & $\times$ & ${\mathbb{R}^{n \times n \times n}}$ & ${\mathbb{R}^{n \times n \times n}}$ & $\times$ \\
        $\varphi _{({t_0},t)}^{i,{p_1}{p_2}}$ & $\times$ & $\times$ & $\times$ & ${\mathbb{R}^{m \times n \times n}}$ & $\times$ & $\times$ & ${\mathbb{R}^{m \times n \times n}}$ \\
        $\phi _{({t_0},t)}^{i,{k_1}{k_2}{k_3}}$ & $\times$ & $\times$ & $\times$ & $\times$ & ${\mathbb{R}^{n \times n \times n \times n}}$ & ${\mathbb{R}^{n \times n \times n \times n}}$ & $\times$ \\
        $\varphi _{({t_0},t)}^{i,{p_1}{p_2}{p_3}}$ & $\times$ & $\times$ & $\times$ & $\times$ & $\times$ & $\times$ & ${\mathbb{R}^{m \times n \times n \times n}}$ \\
        Total & \makecell[c]{$n+n^2$\\(42)} & \makecell[c]{$n^2+n^3+$\\$n$ (256)} & \makecell[c]{$2n+2n^2+$\\ $nm^2$ (90)} & \makecell[c]{$n+n^2+nm^2+$\\$(n+1)m$ (55)} & \makecell[c]{$n+n^2+n^3$\\$+n^4$ (1554)} & \makecell[c]{$nm^2(1+m)$\\ $2n(1+n)$ (96)} & \makecell[c]{$n+n^2+(n+1)m$+\\$n(m^2+m^3)$ (61)} \\ 
        \hline\hline
\end{tabular}
\end{table}

\section{Numerical Examples}
\label{Sec:Numerical Examples}
In this section, several numerical examples are presented to show the advantages of the proposed method. Simulations are performed on a personal laptop with a 12th Gen Intel(R) Core(TM)i5-12500H processor and 16 GB memory.

\subsection{Jupiter Case}
\label{Sec:Jupiter Case}
\begin{figure}[!h]
    \centering
    \subfigure[Nominal orbit]{
        \label{fig:1a}
        \includegraphics[width=0.45\linewidth]{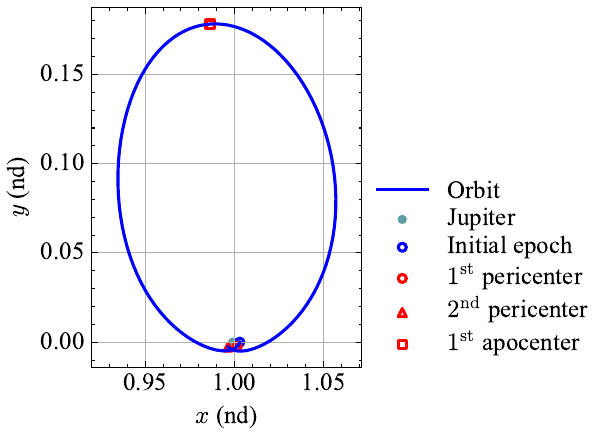}
    }
    \subfigure[Larger plot of \subref{fig:1a}]{
        \label{fig:1b}
        \includegraphics[width=0.45\linewidth]{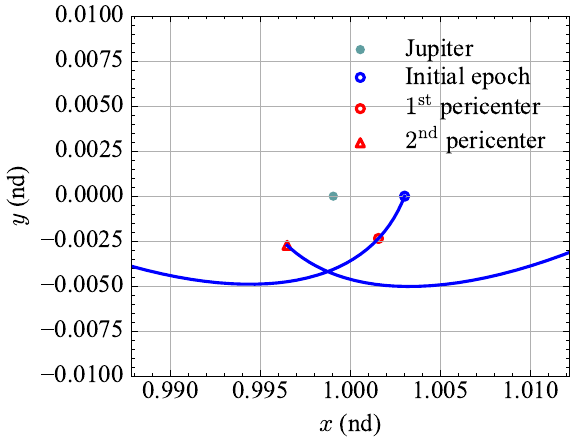}
    }
    \caption{\label{fig:Nominal orbit of the Jupiter case} Nominal orbit of the Jupiter case.}
\end{figure}

In this subsection, the proposed TDSTT is applied to predict the orbital uncertainties of a TC orbit in the Sun-Jupiter three-body system. Let $\boldsymbol{x} = [\boldsymbol{r};\boldsymbol{v}] = {[x,y,z,\dot x,\dot y,\dot z]^T}$ be the nondimensional orbital state of the spacecraft in the Sun-Jupiter rotating frame, with $\boldsymbol{r} = {[x,y,z]^T}$ and $\boldsymbol{v} = {[\dot x,\dot y,\dot z]^T}$ being the nondimensional position and velocity vectors, respectively. The Sun-Jupiter circular-restricted three-body problem (CRTBP) is employed to approximate the high-fidelity dynamics of the Sun-Jupiter system, with the mathematical model formulated as
\begin{equation} \label{eq:CRTBP}
    \left\{ {\begin{array}{*{20}{l}}
        {\ddot x = 2\dot y + x - \frac{{(1 - \mu )(x + \mu )}}{{r_1^3}} - \frac{{\mu (x + \mu  - 1)}}{{r_2^3}}}\\
        {\ddot y =  - 2\dot x + y - \frac{{(1 - \mu )y}}{{r_1^3}} - \frac{{\mu y}}{{r_2^3}}}\\
        {\ddot z =  - \frac{{(1 - \mu )z}}{{r_1^3}} - \frac{{\mu z}}{{r_2^3}}}
    \end{array}} \right. \,,
\end{equation}
where ${r_1} = \sqrt {{{(x + \mu )}^2} + {y^2} + {z^2}}$, ${r_2} = \sqrt {{{(x + \mu  - 1)}^2} + {y^2} + {z^2}}$, and $\mu$ is the nondimensional gravitational constant of the Sun-Jupiter three-body system. The nominal orbit is adopted from Ref. \cite{Li2024MNRAS}. Necessary parameters for the TC orbit, including the nondimensional gravitational constant, initial nondimensional position and velocity vectors, and the initial and final epochs, are listed in Table~\ref{tab:Parameters for the Jupiter case}. The corresponding nominal orbit is plotted in Fig.~\ref{fig:Nominal orbit of the Jupiter case}, with its pericenters and apocenter represented by red markers. As shown in Fig.~\ref{fig:Nominal orbit of the Jupiter case}, the spacecraft is temporarily captured by Jupiter’s gravitational solid influence upon reaching its first pericenter (the red circle in Fig.~\ref{fig:Nominal orbit of the Jupiter case}) rather than directly escaping the Jupiter system. After that, under the combined gravitational effects of the Sun and Jupiter, the spacecraft reaches its apocenter (the red square in Fig.~\ref{fig:Nominal orbit of the Jupiter case}) and returns to the Jupiter system, approaching a second pericenter (the red triangle in Fig.~\ref{fig:Nominal orbit of the Jupiter case}) relative to Jupiter. 

\begin{table}[!h]
    \caption{\label{tab:Parameters for the Jupiter case} Parameters for the Jupiter case}
    \centering
    \begin{tabular}{lrl}
        \hline\hline
        \multicolumn{2}{l}{Parameter}                     & Value                \\ \hline
        \multicolumn{2}{l}{$\mu$}                         & 0.000953886085903286 \\ \hline
        \multirow{3}{*}{Position vector (nd)} & $x$       & 1.00300694584498     \\
                                              & $y$       & 0                    \\
                                              & $z$       & 0                    \\ \hline
        \multirow{3}{*}{Velocity vector (nd)} & $\dot{x}$ & -0.247985627039792   \\
                                              & $\dot{y}$ & -0.646024645202596   \\
                                              & $\dot{z}$ & 0                    \\ \hline
        \multicolumn{2}{l}{Initial epoch $t_0$ (nd)}         & 0                    \\ \hline
        \multicolumn{2}{l}{Final epoch $t_f$ (nd)}           & 3.14815010456319     \\ \hline
        \multicolumn{2}{l}{Epoch of the $1^{\rm{st}}$ pericenter} (nd) & 0.00383531150587122  \\ \hline
        \multicolumn{2}{l}{Epoch of the $1^{\rm{st}}$ apocenter} (nd)  & 1.55033345501027     \\ \hline
        \multicolumn{2}{l}{Epoch of the $2^{\rm{nd}}$ pericenter} (nd) & 3.14815010456319     \\
        \hline\hline
    \end{tabular}
\end{table}

\begin{figure}[!h]
    \centering
    \includegraphics[width=0.45\linewidth]{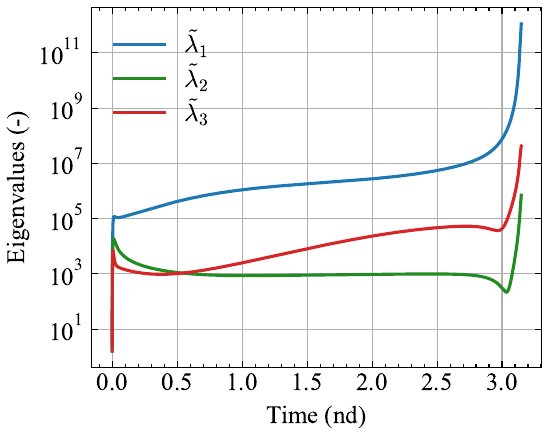}
    \caption{\label{fig:Time histories of the first three eigenvalues of the Jupiter case} Time histories of the first three eigenvalues of the Jupiter case.}
\end{figure}

To begin with, a second-order TDSTT with three sensitive directions (\emph{i.e.}, $m=3$) is propagated. Numerical integration is implemented in a Python environment using a Runge-Kutta (4,5) solver (RK45) \footnote{Using the Python package scipy: \url{https://pypi.org/project/scipy/}}. The small step to start the integration of the TDSTT is 1/100000 of the total propagation period (\emph{i.e.}, $t' = {t_f}/100000$). The time histories of the first three eigenvalues obtained from the TDSTT are shown in Fig.~\ref{fig:Time histories of the first three eigenvalues of the Jupiter case}. The other three eigenvalues are smaller than 1 (\emph{i.e.}, they represent the stable directions) and are not plotted in Fig.~\ref{fig:Time histories of the first three eigenvalues of the Jupiter case}. As shown in Fig.~\ref{fig:Time histories of the first three eigenvalues of the Jupiter case}, the three eigenvalues are labeled as ${\tilde \lambda _1}$, ${\tilde \lambda _2}$, and ${\tilde \lambda _3}$, respectively. Tildes are added to these three eigenvalues as they are not strictly in a descending order. For example, when the propagated arc is smaller than 0.5 nd, ${\tilde \lambda _2}$ is larger than ${\tilde \lambda _3}$ (\emph{i.e.}, ${\lambda _2} = {\tilde \lambda _2}$ and ${\lambda _3} = {\tilde \lambda _3}$), whereas ${\tilde \lambda _2}$ is smaller than ${\tilde \lambda _3}$ when the time is larger than 0.5 nd (\emph{i.e.}, ${\lambda _2} = {\tilde \lambda _3} > {\lambda _3} = {\tilde \lambda _2}$). 
This shows a potential limitation that the TDSTT cannot follow any swap in the magnitude of the eigenvalues. The rank of the eigenvalues at the epoch of the warm start (\emph{i.e.}, $t'$) determines the sensitive directions employed from $t'$ to $t_f$.
In addition, it can be seen from Fig.~\ref{fig:Time histories of the first three eigenvalues of the Jupiter case} that the largest eigenvalue (\emph{i.e.}, ${\lambda _1} = {\tilde \lambda _1}$) is approximately $10^{12}$ when the spacecraft arrives at its second pericenter (\emph{i.e.}, the final epoch $t_f$). This means that the nonlinearity of such a TC orbit is very high, and an initial orbital state deviation can be magnified by $10^{12}$ after the propagation. In this case, a first-order STT (\emph{i.e.}, STM) is inadequate to generate accurate predictions, necessitating a high-order method. Moreover, one can see from Fig.~\ref{fig:Time histories of the first three eigenvalues of the Jupiter case} that the largest eigenvalue (\emph{i.e.}, approximately $10^{12}$) is at least four orders of magnitude larger than others (the second largest eigenvalue is approximately $10^8$), indicating that the STT terms along the direction associated with the largest eigenvalue (\emph{i.e.}, ${\boldsymbol{\xi }_1}$) dominates over the terms along other directions.

\begin{figure}[!h]
    \centering
    \includegraphics[width=0.45\linewidth]{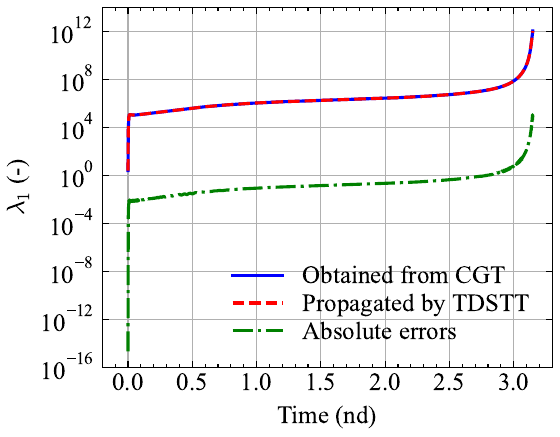}
    \caption{\label{fig:Profiles of the first eigenvalue of the Jupiter case} Profiles of the first eigenvalue of the Jupiter case.}
\end{figure}

Figure~\ref{fig:Profiles of the first eigenvalue of the Jupiter case} gives the profiles of the first eigenvalue (\emph{i.e.}, the largest eigenvalue). The magnitudes of $\lambda_1$ obtained from the CGT (\emph{i.e.}, using Eq.~\eqref{eq:CGT}) and the propagated by the TDSTT (\emph{i.e.}, using Eq.~\eqref{eq:eigenvalue ODE}) are represented by the blue solid and red dashed lines, respectively, and their difference (\emph{i.e.}, the absolute errors) is plotted using a green line. One can see from Fig.~\ref{fig:Profiles of the first eigenvalue of the Jupiter case} that the absolute errors are several orders of magnitude smaller than the absolute values of CGT’s and TDSTT’s eigenvalues. For example, at the final epoch $t_f$, the absolute error is approximately $10^5$, which is only $\frac{1}{10^5}$ (with a relative error of 0.001\%) of the absolute eigenvalue magnitude. The minor discrepancies between the CGT’s and TDSTT’s eigenvalues are caused by the numerical integration errors.

\begin{figure}[!h]
    \centering
    \includegraphics[width=0.45\linewidth]{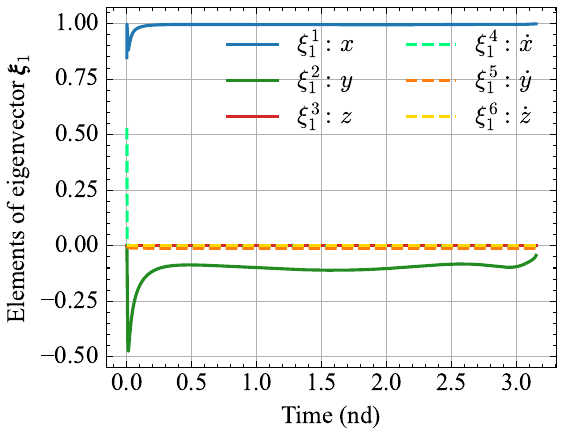}
    \caption{\label{fig:Time histories of the eigenvector elements of the Jupiter case} Time histories of the eigenvector elements of the Jupiter case.}
\end{figure}

\begin{figure}[!h]
    \centering
    \includegraphics[width=0.45\linewidth]{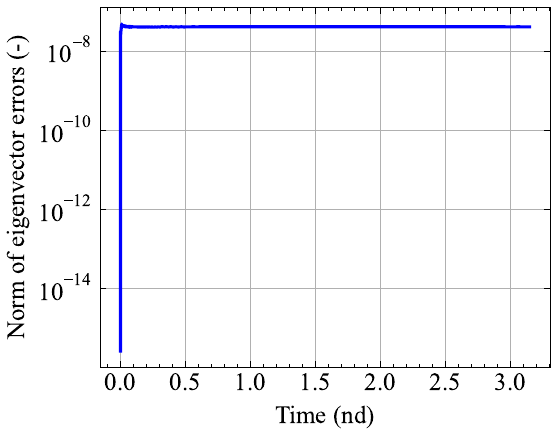}
    \caption{\label{fig:Eigenvector error profile of the Jupiter case} Eigenvector error profile of the Jupiter case.}
\end{figure}

Table~\ref{tab:Elements of the first three eigenvectors in the Jupiter case} gives the eigenvector elements (propagated by TDSTT) at the epoch $t'$ (\emph{i.e.}, the epoch for the warm start) and the final epoch $t_f$. Again, tildes are added to these three eigenvectors as they are associated with the eigenvalues shown in Fig.~\ref{fig:Time histories of the first three eigenvalues of the Jupiter case}. When the propagated time is less than 0.5 nd, one has ${\boldsymbol{\xi }_1} = {\boldsymbol{\tilde \xi }_1}$, ${\boldsymbol{\xi }_2} = {\boldsymbol{\tilde \xi }_2}$, and ${\boldsymbol{\xi }_3} = {\boldsymbol{\tilde \xi }_3}$, whereas after $t = 0.5\,{\rm{nd}}$, one has ${\boldsymbol{\xi }_1} = {\boldsymbol{\tilde \xi }_1}$, ${\boldsymbol{\xi }_2} = {\boldsymbol{\tilde \xi }_3}$, and ${\boldsymbol{\xi }_3} = {\boldsymbol{\tilde \xi }_2}$. The time histories of the six elements of the first eigenvalue are shown in Fig.~\ref{fig:Time histories of the eigenvector elements of the Jupiter case}. One can see from Table~\ref{tab:Elements of the first three eigenvectors in the Jupiter case} and Fig.~\ref{fig:Time histories of the eigenvector elements of the Jupiter case} that the eigenvectors change as time evolves, and that's why the DSTT is only valid at the predefined final epoch $t_f$. For example, the first eigenvector (corresponding to the largest eigenvalue $\tilde{\lambda}_{1}$), ${\boldsymbol{\xi }_1}$, changes from $[0.84,0,0,0.53,0,0]^T$ (elements with absolute values smaller than $10^{-2}$ are replaced with 0) to $[0.99,0.04,0,0,0,0]^T$. To test the numerical integration accuracy of the eigenvectors, Fig.~\ref{fig:Eigenvector error profile of the Jupiter case} presents the eigenvector errors of the first (largest) eigenvector ${\boldsymbol{\xi }_1}$. The eigenvector error is defined as the L2 norm of the true eigenvector subtracted by the propagated eigenvector. As shown in Fig.~\ref{fig:Eigenvector error profile of the Jupiter case}, the errors are less than $10^{-7}$, which is acceptable as the magnitude of the true eigenvectors is 1.

\begin{table}[!h]
    \caption{\label{tab:Elements of the first three eigenvectors in the Jupiter case} Elements of the first three eigenvectors in the Jupiter case}
    \centering
    \begin{tabular}{lrcccccc}
        \hline\hline
        \multicolumn{2}{l}{Eigenvector} & $1^{\rm{st}}$ element & $2^{\rm{nd}}$ element & $3^{\rm{rd}}$ element & $4^{\rm{th}}$ element & $5^{\rm{th}}$ element & $6^{\rm{th}}$ element \\ \hline
        \multirow{2}{*}{$\boldsymbol{\tilde \xi }_1$} & $t'$ & 0.8474 & $-2.2 \times 10^{-3}$ & $2.4 \times 10^{-19}$ & 0.5309 & $-1.3 \times 10^{-3}$ & $-7.3 \times 10^{-20}$ \\
        & $t_f$ & 0.9989 & -0.0436 & $8.1 \times 10^{-23}$ & $-3.6 \times 10^{-3}$ & -0.0107 & $4.0 \times 10^{-26}$ \\ \hline
        \multirow{2}{*}{$\boldsymbol{\tilde \xi }_2$} & $t'$ & $-2.0 \times 10^{-3}$ & -0.7859 & $-2.3 \times 10^{-12}$ & $1.6 \times 10^{-3}$ & 0.6182 & $1.8 \times 10^{-12}$  \\
        & $t_f$ & $5.0 \times 10^{-18}$ & $-4.5 \times 10^{-17}$ & 0.9999 & $9.8 \times 10^{-17}$ & $5.9 \times 10^{-16}$ & $2.4 \times 10^{-3}$ \\ \hline
        \multirow{2}{*}{$\boldsymbol{\tilde \xi }_3$} & $t'$ & $4.1 \times 10^{-15}$ & $1.6 \times 10^{-12}$ & -0.7859 & $-3.2 \times 10^{-15}$ & $-1.2 \times 10^{-12}$ & 0.6182 \\
        & $t_f$ & -0.0436 & -0.9989 & $4.8 \times 10^{-17}$ & 0.0105 & $-3.1 \times 10^{-3}$ & $2.4 \times 10^{-20}$ \\ \hline\hline
    \end{tabular}
\end{table}

The results in Fig.~\ref{fig:Profiles of the first eigenvalue of the Jupiter case} and Fig.~\ref{fig:Eigenvector error profile of the Jupiter case} show that the CGT’s eigenvalues and eigenvectors can be correctly propagated using the TDSTT. Then, instead of propagating one single sample, a set of neighboring orbits is propagated over time to evaluate the accuracy performance of the developed TDSTT. A total of 30 samples are considered, with their initial states forming a circumference with a radius of $1.3 \times 10^{-7}$ (approximately 100 km) lying on the \emph{x}-\emph{y} plane and centered at the initial state of the nominal orbit (\emph{i.e.}, $\boldsymbol{x}_{0}$); thus, the initial state deviations are expressed as
\begin{equation} \label{eq:dx0}
    \delta {\boldsymbol{x}_0} = 1.3 \times {10^{-7}} \times \left [\sin {\alpha _k}, \cos {\alpha _k}, 0, 0, 0, 0 \right ]^{T} \,,
\end{equation}
where ${\alpha _k} = \frac{{k\pi }}{{15}}$ ($k \in \{ 1, \cdots ,30\} $). These 30 neighboring orbits and the nominal orbit are propagated using the RK45 solver, and the corresponding state deviations are predicted using the full STTs up to the third order and the TDSTTs at the same order. For the TDSTTs, one sensitive direction is employed (\emph{i.e.}, $m=1$). The projections of the state deviations (at the final epoch $t_f$) on the \emph{x}-\emph{y} plane are shown in Fig.~\ref{fig:Projections of orbital state deviations predicted by the STM, 2nd-order STT, and 2nd-order TDSTT} and Fig.~\ref{fig:Projections of orbital state deviations predicted by the STM, 3rd-order STT, and 3rd-order TDSTT}. The propagated state deviations (can be considered true results) are plotted using green markers, and the results predicted by the STM, high-order (full) STTs, and TDSTTs are represented by red, orange, and yellow markers, respectively. In addition, the prediction errors of the full STTs and TDSTTs are presented in the right subfigures in Fig.~\ref{fig:Projections of orbital state deviations predicted by the STM, 2nd-order STT, and 2nd-order TDSTT} and Fig.~\ref{fig:Projections of orbital state deviations predicted by the STM, 3rd-order STT, and 3rd-order TDSTT}. The state deviations along the \emph{z}-axis are not shown in Fig.~\ref{fig:Projections of orbital state deviations predicted by the STM, 2nd-order STT, and 2nd-order TDSTT} and Fig.~\ref{fig:Projections of orbital state deviations predicted by the STM, 3rd-order STT, and 3rd-order TDSTT} as they are several orders of magnitude smaller than the deviations along the other two axes.

\begin{figure}[!h]
    \centering
    \includegraphics[width=0.95\linewidth]{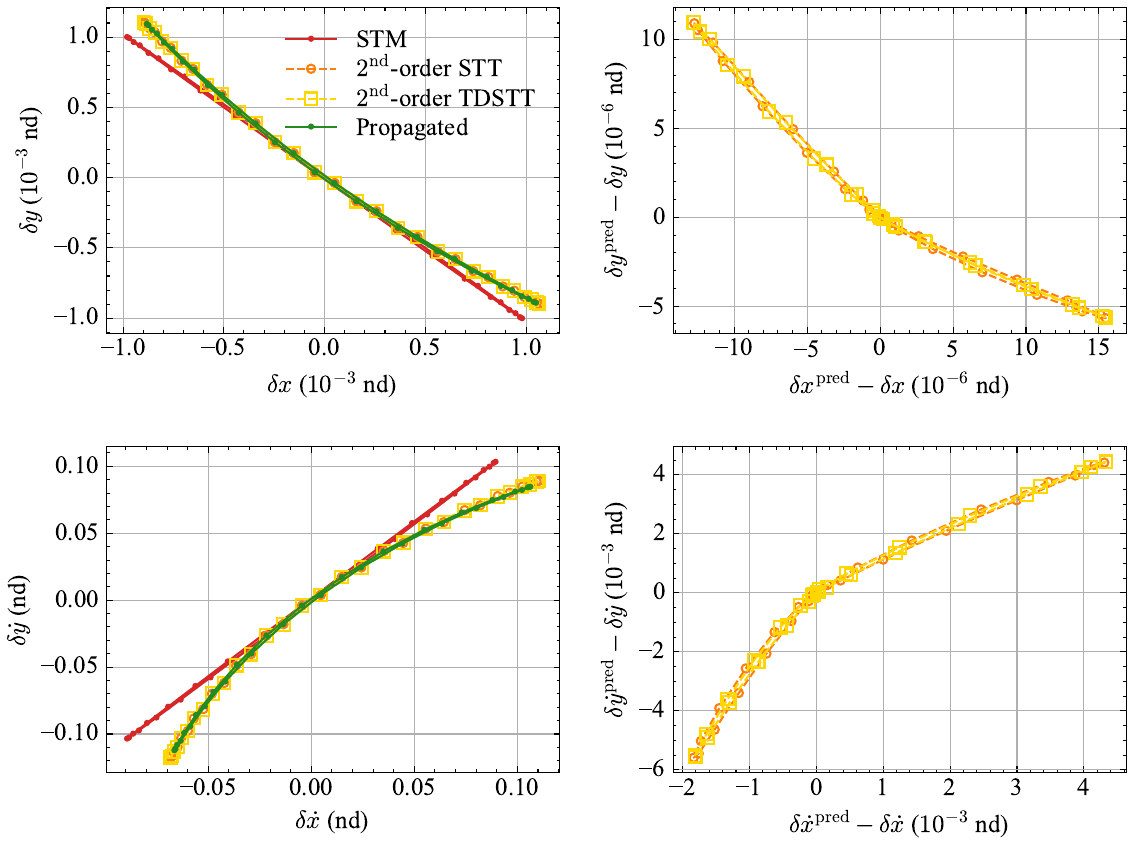}
    \caption{\label{fig:Projections of orbital state deviations predicted by the STM, 2nd-order STT, and 2nd-order TDSTT} Projections of orbital state deviations predicted by the STM, $2^{\rm{nd}}$-order STT, and $2^{\rm{nd}}$-order TDSTT.}
\end{figure}

\begin{figure}[!h]
    \centering
    \includegraphics[width=0.95\linewidth]{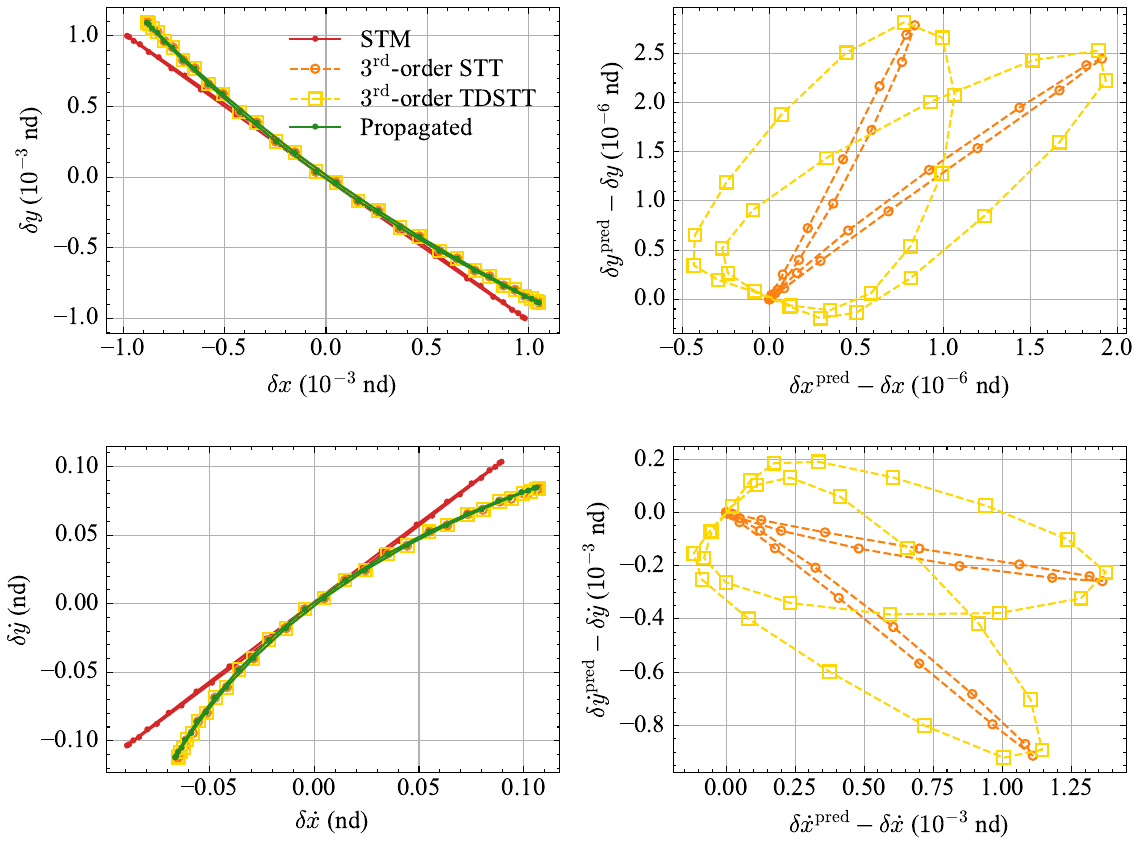}
    \caption{\label{fig:Projections of orbital state deviations predicted by the STM, 3rd-order STT, and 3rd-order TDSTT} Projections of orbital state deviations predicted by the STM, $3^{\rm{rd}}$-order STT, and $3^{\rm{rd}}$-order TDSTT.}
\end{figure}

As shown in Fig.~\ref{fig:Projections of orbital state deviations predicted by the STM, 2nd-order STT, and 2nd-order TDSTT} and Fig.~\ref{fig:Projections of orbital state deviations predicted by the STM, 3rd-order STT, and 3rd-order TDSTT}, affected by the highly nonlinear effects of the Sun-Jupiter system, the state deviations evolve nonlinearly; thus, a linear method (\emph{i.e.}, the STM) is insufficient to predict state deviations in this case. The green (true), orange (STT), and yellow (TDSTT) markers overlap with each other, indicating that both the high-order STTs and the developed TDSTTs can capture the nonlinearity and produce more accurate state deviation predictions. The prediction errors of the full STTs and TDSTTs can be reduced if higher-order terms are included. The second-order STT and TDSTT have position errors of approximately $10^{-5}$, whereas the errors reduce to $2.5 \times 10^{-6}$ using the third-order STT and TDSTT. In addition, the prediction errors of the TDSTT are slightly larger than those of the full STT with the same order. This is expected because the TDSTT is an approximation of the full STT. Like the DSTT, the TDSTT neglects the high-order terms of the full STT along the insensitive directions.

\begin{figure}[!h]
    \centering
    \includegraphics[width=0.95\linewidth]{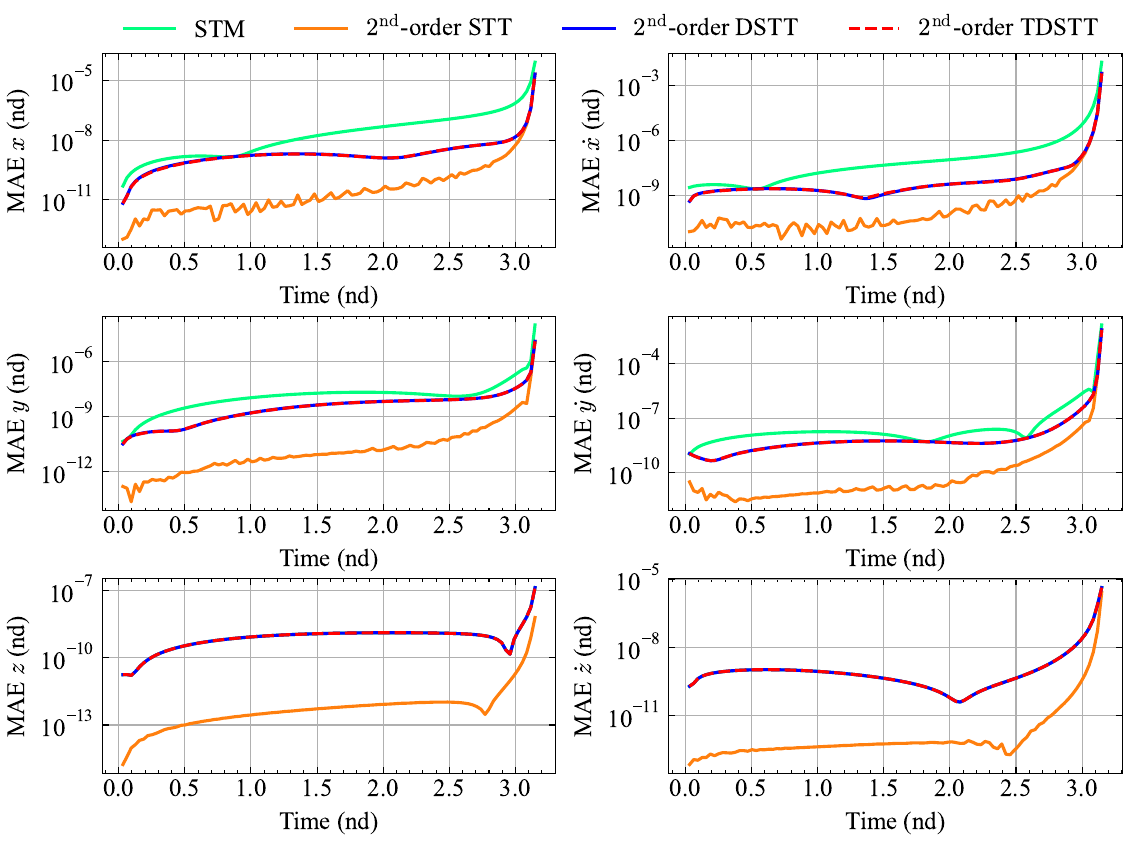}
    \caption{\label{fig:Profiles of the mean absolute errors of different methods} Profiles of the mean absolute errors of different methods.}
\end{figure}

A Monte Carlo (MC) simulation is implemented to investigate the developed TDSTT’s performance further. Ten thousand MC runs are performed. In each MC run, the initial state is randomly generated from a normal distribution ${\cal N}({\boldsymbol{x}_0},{\boldsymbol{P}_0})$, with $\boldsymbol{x}_{0}$ being the initial state of the nominal orbit and $\boldsymbol{P}_{0} \in \mathbb{R}^{6 \times 6}$ being the covariance matrix. The position and velocity standard deviations (STDs) of the normal distribution are set as $1.3 \times 10^{-7}$ (approximately 100 km) per axis and $7.6 \times 10^{-7}$ (approximately 10 mm/s) per axis, respectively; thus, the covariance matrix $\boldsymbol{P}_{0}$ can be written as
\begin{equation} \label{eq:P0}
    {\boldsymbol{P}_0} = \left[ {\begin{array}{*{20}{c}}
        {{{1.3}^2} \times {{10}^{ - 14}}{\boldsymbol{I}_3}} & {{\mathbf{0}_{3 \times 3}}}\\
        {{\mathbf{0}_{3 \times 3}}} & {{{7.6}^2} \times {{10}^{ - 14}}{\boldsymbol{I}_3}}
    \end{array}} \right] \,,
\end{equation}
where $\boldsymbol{I}_{3}$ is a 3-dimensional identified matrix, and $\mathbf{0}_{3 \times 3}$ is a zero matrix of size $3 \times 3$. The STM, second-order (full) STT, second-order DSTT, and second-order TDSTT are employed to predict the state deviations of the neighboring orbits relative to the nominal orbit. Here, the number of the sensitive direction is set as 1 for both the DSTT and TDSTT (\emph{i.e.}, $m=1$). The DSTT terms are computed using the indirect way, \emph{i.e.}, by projecting the full STT terms in the direction of the CGT’s eigenvector associated with the largest eigenvalue (\emph{i.e.}, using Eqs.~\eqref{eq:DSTT1}-\eqref{eq:DSTT2}). The mean absolute errors (MAEs) of the MC simulation are evaluated at 1000 epochs evenly distributed from the initial epoch $t_0$ to the predefined final epoch $t_f$. By connecting these MAE points, the corresponding MAE curves are depicted in Fig.~\ref{fig:Profiles of the mean absolute errors of different methods}. Recall that if one wants to analyze the evolution of orbital uncertainties (like Fig.~\ref{fig:Profiles of the mean absolute errors of different methods}), both direct and indirect ways of generating DSTTs cannot reduce computational costs. The indirect way requires the additional computation of the eigenvalue-eigenvector pairs of the CGT, whereas the direct way requires repeatedly integrating the DSTTs for analysis at different epochs (1000 times for the results in Fig.~\ref{fig:Profiles of the mean absolute errors of different methods}).

As depicted in Fig.~\ref{fig:Profiles of the mean absolute errors of different methods}, the TDSTT has the same level of accuracy as the DSTT. The STM has the largest MAEs, and the full STT is more accurate than the other three methods. The MAEs of the DSTT and TDSTT are close to those of the STM in the beginning, and they converge to the MAE curves of the exact second-order solution (\emph{i.e.}, the second-order full STT) as time advances. This is because, in the beginning, the largest eigenvalue is only one order of magnitude larger than the second largest eigenvalue (as shown in Fig.~\ref{fig:Time histories of the first three eigenvalues of the Jupiter case}), indicating that the effects of the terms along the eigenvector direction associated with the second largest eigenvalue are comparable. However, as time advances, the discrepancies between the first eigenvalue and other eigenvalues become larger (again, see Fig.~\ref{fig:Time histories of the first three eigenvalues of the Jupiter case}), and the terms along the direction of the first eigenvector (associated with the largest eigenvalue) tend to dominate the propagation of the state deviation. In this case, the improvements of the DSTT and TDSTT against the linear solution (\emph{i.e.}, STM) grow over time. It is also interesting to see from Fig.~\ref{fig:Profiles of the mean absolute errors of different methods} that, for the position and velocity states along the \emph{z}-axis, the MAE curves of the STM, DSTT, and TDSTT overlap. This means that the DSTT and TDSTT don’t provide improvements in predicting the state deviations along the \emph{z}-axis against the linear solution. Our explanation is given as follows. As shown in Table~\ref{tab:Elements of the first three eigenvectors in the Jupiter case} and Fig.~\ref{fig:Time histories of the eigenvector elements of the Jupiter case}, the elements of the first eigenvector along the \emph{z}-axis (\emph{i.e.}, $\xi _1^3$ and $\xi _1^6$) are almost zero (approximately $10^{-20}$); thus, the high-order terms along the \emph{z}-axis are neglected by both the DSTT and TDSTT.

\begin{figure}[!h]
    \centering
    \subfigure[RE of $x(t_f)$]{
        \label{fig:9a}
        \includegraphics[width=0.75\linewidth]{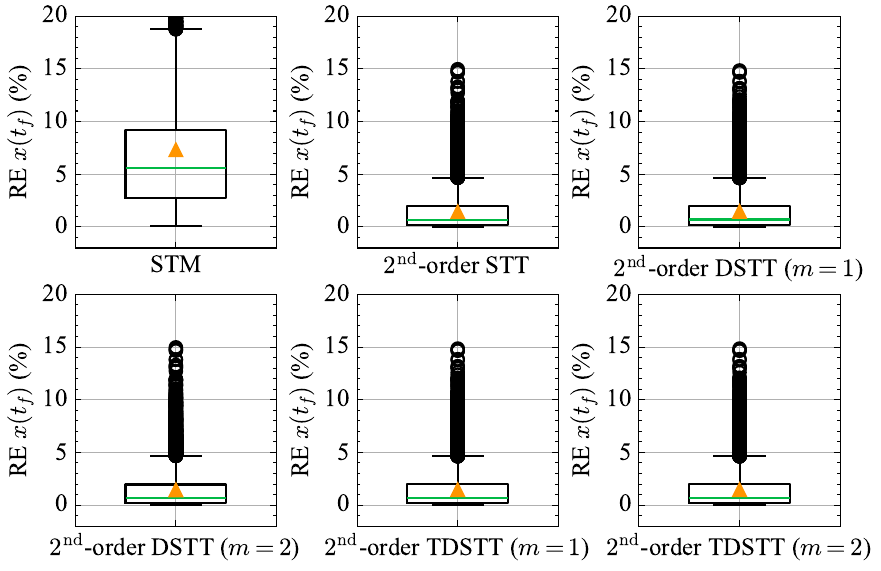}
    }
    \subfigure[RE of $y(t_f)$]{
        \label{fig:9b}
        \includegraphics[width=0.75\linewidth]{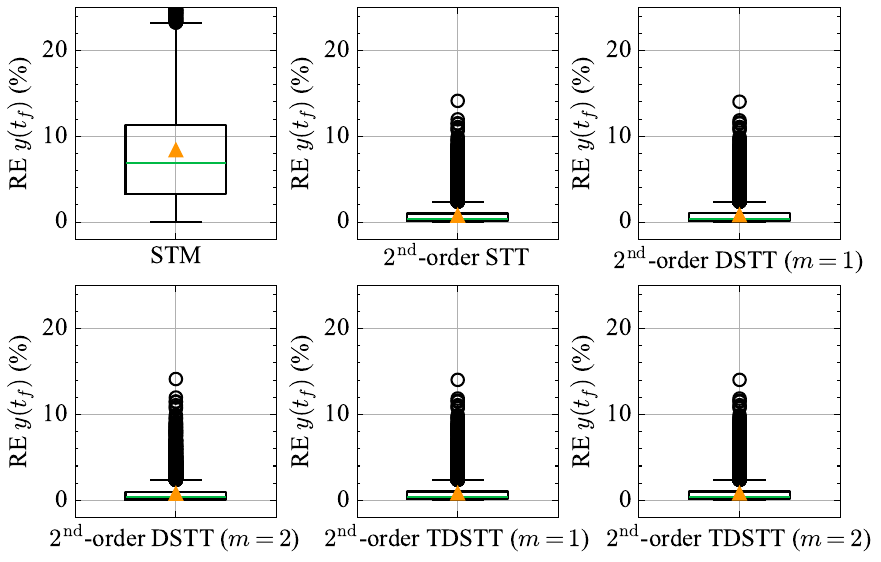}
    }
    \caption{\label{fig:Boxplots of the relative errors in the Jupiter case} Boxplots of the relative errors in the Jupiter case.}
\end{figure}

Figure~\ref{fig:Boxplots of the relative errors in the Jupiter case} shows the distributions of the predicted relative errors (REs) of 10000 MC suns at the final epoch $t_f$. In Fig.~\ref{fig:Boxplots of the relative errors in the Jupiter case}, the RE distributions are represented graphically using boxplots (also known as box-and-whisker plots). A boxplot usually displays the minimum, first quartile, median, third quartile, and maximum of the data (\emph{i.e.}, REs). The box spans the interquartile range, which represents the middle 50\% of the data. The whiskers extend from the edges of the box to show the range of the data. Outliers (black circles in Fig.~\ref{fig:Boxplots of the relative errors in the Jupiter case}) are also shown as individual points beyond the whiskers, and the median and the mean are highlighted using green lines and orange triangles, respectively. One can see from Fig.~\ref{fig:Boxplots of the relative errors in the Jupiter case} that, for the Jupiter case, the TDSTT can present nearly the same level of accuracy as the full STT and the DSTT. The mean relative errors (MREs) of the TDSTT, full STT, and DSTT are approximately 1\% (see the orange triangles in Fig.~\ref{fig:Boxplots of the relative errors in the Jupiter case}), while the linear solution (\emph{i.e.}, STM) has MREs larger than 7.5\%. In most MC runs, the REs of the TDSTT, full STT, and DSTT are less than 5\%; however, in more than 25\% of the MC runs, the STM has REs larger than 10\%.

Table~\ref{tab:Mean absolute errors at the final epoch of the Jupiter case} lists the MAE results of different methods at the final epoch. Using the linear solution (\emph{i.e.}, the STM), the MAEs for the position and velocity are approximately $1 \times 10^{-4}$ and $1 \times 10^{-2}$, respectively. The MAEs can be reduced by around one order of magnitude if a second-order method is employed. The TDSTT has slightly larger MAEs than the DSTT and STT with the corresponding order. For example, the MAE along the \emph{x}-axis of the third-order TDSTT ($m=2$) is $3.94 \times 10^{-6}$, which is 4.72\% larger than that of the third-order DSTT (its MAE is $3.76 \times 10^{-6}$). This is because the TDSTT is an approximation of the DSTT. Note that approximations are employed in Eqs.~\eqref{eq:DSTT-STT1}-\eqref{eq:DSTT-STT} when deriving the differential equations of the TDSTT. It is also interesting to see from Table~\ref{tab:Mean absolute errors at the final epoch of the Jupiter case} that when two sensitive directions are used (\emph{i.e.}, $m=2$), the TDSTT provides more accurate predictions for the state deviations along the \emph{z}-axis than the DSTT. The MAEs of the position deviation along the \emph{z}-axis of the second-order TDSTT ($m=2$) and second-order DSTT ($m=2$) are $6.25 \times 10^{-9}$ and $1.35 \times 10^{-7}$, respectively. Our explanation is provided as follows. The DSTT ranks the eigenvalues and selects the eigenvectors (\emph{i.e.}, the sensitive directions) according to the CGT. As shown in Fig.~\ref{fig:Time histories of the first three eigenvalues of the Jupiter case}, at the beginning, the magnitude of ${\tilde \lambda _2}$ is larger than the magnitude of ${\tilde \lambda _3}$, while as time advances, ${\tilde \lambda _3}$ comes from behind and finally is larger than ${\tilde \lambda _2}$. Thus, for the DSTT ($m=2$), the eigenvectors associated with ${\tilde \lambda _1}$ and ${\tilde \lambda _3}$ are employed to construct the linear transformation matrix $\boldsymbol{R}$. Unlike the DSTT, the TDSTT selects eigenvectors in the warm start step, and whereafter, these selected eigenvectors are integrated along the nominal orbit. Hence, the eigenvectors associated with ${\tilde \lambda _1}$ and ${\tilde \lambda _2}$ are chosen as sensitive directions for the TDSTT. As shown in Table~\ref{tab:Elements of the first three eigenvectors in the Jupiter case}, although the direction along ${\boldsymbol{\tilde \xi }_3}$ is more sensitive than the direction along ${\boldsymbol{\tilde \xi }_2}$ (from a view of CGT), it doesn’t contain the elements along the \emph{z}-axis. The DSTT ($m=2$) aligns the high-order terms along the directions of ${\boldsymbol{\tilde \xi }_1}$ and ${\boldsymbol{\tilde \xi }_3}$, preserving no terms along the \emph{z}-axis. However, the TDSTT ($m=2$) takes ${\boldsymbol{\tilde \xi }_1}$ and ${\boldsymbol{\tilde \xi }_2}$ as sensitive directions, and thus, its solution includes high-order terms along the \emph{z}-axis.

\begin{table}[!h]
    \caption{\label{tab:Mean absolute errors at the final epoch of the Jupiter case} Mean absolute errors at the final epoch of the Jupiter case}
    \centering
    \begin{tabular}{llcccccc}
        \hline\hline
        \multirow{2}{*}{Order} & \multirow{2}{*}{Method} & \multicolumn{3}{c}{Position (nd)} & \multicolumn{3}{c}{Velocity (nd)} \\ \cline{3-8}
        & & $x$ & $y$ & $z$ & $\dot{x}$ & $\dot{y}$ & $\dot{z}$ \\ \hline
        1 & STM & 8.24$\times 10^{-5}$ & 9.84$\times 10^{-5}$ & 1.35$\times 10^{-7}$ & 1.82$\times 10^{-2}$ & 1.29$\times 10^{-2}$ & 4.21$\times 10^{-6}$ \\ \hline
        \multirow{5}{*}{2} & STT & 2.08$\times 10^{-5}$ & 1.25$\times 10^{-5}$ & 6.24$\times 10^{-9}$ & 4.49$\times 10^{-3}$ & 6.98$\times 10^{-3}$ & 2.54$\times 10^{-6}$ \\
        & DSTT ($m=1$) & 2.08$\times 10^{-5}$ & 1.25$\times 10^{-5}$ & 1.35$\times 10^{-7}$ & 4.50$\times 10^{-3}$ & 6.99$\times 10^{-3}$ & 4.21$\times 10^{-6}$ \\
        & TDSTT ($m=1$) & 2.08$\times 10^{-5}$ & 1.25$\times 10^{-5}$ & 1.35$\times 10^{-7}$ & 4.50$\times 10^{-3}$ & 6.99$\times 10^{-3}$ & 4.21$\times 10^{-6}$ \\
        & DSTT ($m=2$) & 2.08$\times 10^{-5}$ & 1.25$\times 10^{-5}$ & 1.35$\times 10^{-7}$ & 4.49$\times 10^{-3}$ & 6.98$\times 10^{-3}$ & 4.21$\times 10^{-6}$ \\
        & TDSTT ($m=2$) & 2.08$\times 10^{-5}$ & 1.25$\times 10^{-5}$ & 6.25$\times 10^{-9}$ & 4.50$\times 10^{-3}$ & 6.99$\times 10^{-3}$ & 2.56$\times 10^{-6}$ \\ \hline
        \multirow{5}{*}{3} & STT & 3.76$\times 10^{-6}$ & 6.91$\times 10^{-6}$ & 2.64$\times 10^{-9}$ & 3.06$\times 10^{-3}$ & 1.75$\times 10^{-3}$ & 4.90$\times 10^{-7}$ \\
        & DSTT ($m=1$) & 3.92$\times 10^{-6}$ & 7.01$\times 10^{-6}$ & 1.35$\times 10^{-7}$ & 3.09$\times 10^{-3}$ & 1.84$\times 10^{-3}$ & 4.21$\times 10^{-6}$ \\
        & TDSTT ($m=1$) & 3.93$\times 10^{-6}$ & 7.02$\times 10^{-6}$ & 1.35$\times 10^{-7}$ & 3.09$\times 10^{-3}$ & 1.84$\times 10^{-3}$ & 4.21$\times 10^{-6}$ \\
        & DSTT ($m=2$) & 3.76$\times 10^{-6}$ & 6.93$\times 10^{-6}$ & 1.35$\times 10^{-7}$ & 3.06$\times 10^{-3}$ & 1.75$\times 10^{-3}$ & 4.21$\times 10^{-6}$ \\
        & TDSTT ($m=2$) & 3.94$\times 10^{-6}$ & 7.01$\times 10^{-6}$ & 2.66$\times 10^{-9}$ & 3.09$\times 10^{-3}$ & 1.84$\times 10^{-3}$ & 5.76$\times 10^{-7}$ \\ 
        \hline\hline
    \end{tabular}
\end{table}

Table~\ref{tab:Number of elements to be integrated for different methods} shows the number of elements to be integrated using different methods. The number of elements required in the first- (STM), second-, and third-order full STTs are 42 ($6+6^2$), 256 ($6+6^2+6^3$), and 1554 ($6+6^2+6^3+6^4$), respectively. The developed TDSTT requires significantly fewer elements than the full STT of the same order. For example, if $m=1$, the second-order TDSTT can reduce 78.5\% of the elements required for the second-order full STT. When the order increases to 3, the improvement grows to 96.1\%. Table~\ref{tab:Computational costs for computing the high-order terms in the Jupiter case} and Table~\ref{tab:Computational costs for predicting state deviations in the Jupiter case} provide the computational costs for computing the high-order terms and predicting the state deviations using different methods. As shown in Table~\ref{tab:Computational costs for computing the high-order terms in the Jupiter case}, the TDSTT requires a higher computational burden than the linear solution (\emph{i.e.}, STM); however, it ($m=2$) can be 60.39\% and 94.14\% faster than the full STT if the terms up to the second and third orders are required, respectively. Note that the computational costs of the DSTT in Table~\ref{tab:Computational costs for computing the high-order terms in the Jupiter case} are evaluated based on one integration from the initial epoch $t_0$ to the predefined final epoch $t_f$. If analysis is required at the predefined final epoch only, the computational costs of numerical integrations of the TDSTT and DSTT are close. Although the DSTT needs to integrate the STM twice, it is very fast to obtain the STM (only 0.32 s, as shown in Table~\ref{tab:Computational costs for computing the high-order terms in the Jupiter case}), posing slight influences on the computational burden. However, if time series analysis is required (like the results in Fig.~\ref{fig:Profiles of the mean absolute errors of different methods}), the DSTT’s computational costs seriously increase. In this case, the computational cost of a second-order DSTT ($m$=2) is 735.86 s (for one thousand epochs in Fig.~\ref{fig:Profiles of the mean absolute errors of different methods}), which is two orders of magnitude larger than that of the TDSTT. In addition, one can see from Table~\ref{tab:Computational costs for predicting state deviations in the Jupiter case} that the proposed TDSTT can also reduce computational costs when applied to predict orbit deviations. Its algorithm complexity grows polynomially as the order increases, making it promising for onboard applications. The developed TDSTT can provide almost the same level of accuracy as the full STT while bringing significantly less computational burden.

\begin{table}[!h]
    \caption{\label{tab:Number of elements to be integrated for different methods} Number of elements to be integrated for different methods}
    \centering
    \begin{tabular}{llc}
        \hline\hline
        Order              & Method      & Number \\ \hline
        1                  & STM         & 42     \\ \hline
        \multirow{5}{*}{2} & STT         & 256    \\
                           & DSTT ($m=1$) & 90 \\
                           & TDSTT ($m=1$) & 55    \\
                           & DSTT ($m=2$) & 108 \\
                           & TDSTT ($m=2$) & 80    \\ \hline
        \multirow{5}{*}{3} & STT         & 1554    \\
                           & DSTT ($m=1$) & 96 \\
                           & TDSTT ($m=1$) & 61    \\
                           & DSTT ($m=2$) & 156 \\
                           & TDSTT ($m=2$) & 128   \\ \hline\hline 
    \end{tabular}
\end{table}

\begin{table}[!h]
    \caption{\label{tab:Computational costs for computing the high-order terms in the Jupiter case} Computational costs for computing the high-order terms in the Jupiter case}
    \centering
    \begin{tabular}{llccc}
        \hline\hline
        \multirow{2}{*}{Order} & \multirow{2}{*}{Method} & \multicolumn{3}{c}{Computational cost (s)} \\ \cline{3-5}
        & & Warm start & Numerical integration & Total \\ \hline
        1 & STM & 0 & 0.3229 & 0.3229 \\ \hline
        \multirow{5}{*}{2} & STT & 0 & 9.7698 & 9.7698 \\
        & DSTT ($m$=1) & 0 & 2.5335 & 2.5335 \\
        & TDSTT ($m$=1) & 0.1773 & 2.2733 & 2.4506 \\
        & DSTT ($m$=2) & 0 & 4.2244 & 4.2244 \\
        & TDSTT ($m$=2) & 0.2093 & 3.6599 & 3.8691 \\ \hline
        \multirow{5}{*}{3} & STT & 0 & 837.4347 & 837.4347 \\
        & DSTT ($m$=1) & 0 & 10.3863 & 10.3863 \\
        & TDSTT ($m$=1) & 15.9723 & 7.4066 & 23.3789 \\
        & DSTT ($m$=2) & 0 & 34.1267 & 34.1267 \\
        & TDSTT ($m$=2) & 15.8685 & 33.1473 & 49.0158 \\ \hline\hline
    \end{tabular}
\end{table}

\begin{table}[!h]
    \caption{\label{tab:Computational costs for predicting state deviations in the Jupiter case} Computational costs for predicting state deviations in the Jupiter case}
    \centering
    \begin{tabular}{llcccc}
        \hline\hline
        \multirow{2}{*}{Order} & \multirow{2}{*}{Method} & \multirow{2}{*}{Algorithm complexity} & \multicolumn{3}{c}{Computational cost (s)} \\ \cline{4-6}
        & & & Averaged & Maximal & Minimal \\ \hline
        1 & STM & $\rm{O}(n^2)$ & 0.0777 & 1.6883 & 0.0089 \\ \hline
        \multirow{5}{*}{2} & STT & $O(n^2+n^3)$ & 0.4117 & 2.8177 & 0.1064 \\
        & DSTT ($m=1$) & $O(n^2+n)$ & 0.0753 & 1.3133 & 0.0191 \\
        & TDSTT ($m=1$) & $O(n^2+n)$ & 0.0861 & 4.7746 & 0.0189 \\
        & DSTT ($m=2$) & $O(n^2+4n)$ & 0.1242 & 1.4020 & 0.0304 \\ 
        & TDSTT ($m=2$) & $O(n^2+4n)$ & 0.1313 & 1.5254 & 0.0325 \\ 
        \hline
        \multirow{5}{*}{3} & STT & $O(n^2+n^3+n^4)$ & 2.9550 & 14.2127 & 0.8054 \\
        & DSTT ($m=1$) & $O(n^2+2n)$ & 0.0981 & 1.0610 & 0.0278  \\
        & TDSTT ($m=1$) & $O(n^2+2n)$ & 0.1113 & 2.6010 & 0.0277 \\
        & DSTT ($m=2$) & $O(n^2+12n)$ & 0.2594 & 1.8721 & 0.0735  \\
        & TDSTT ($m=2$) & $O(n^2+12n)$ & 0.2715 & 2.2649 & 0.0785 \\ \hline\hline
    \end{tabular}
\end{table}

\subsection{Cislunar Case}
\label{Sec:Cislunar Case}
\begin{figure}[!h]
    \centering
    \includegraphics[width=0.7\linewidth]{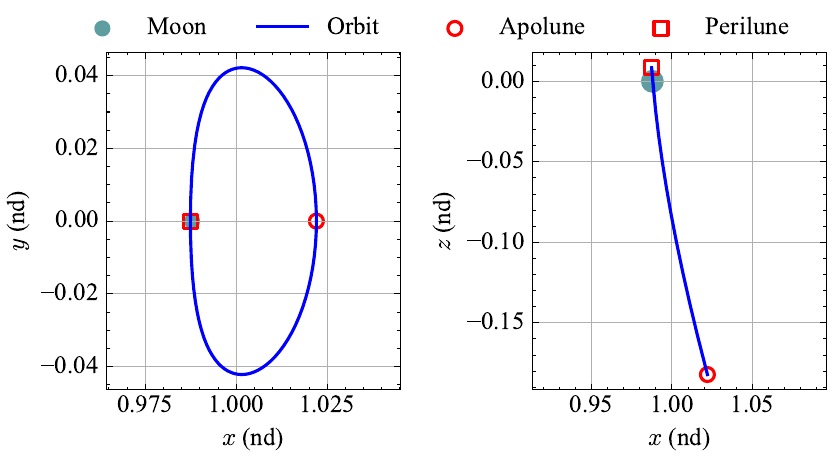}
    \caption{\label{fig:Nominal orbit of the cislunar case} Nominal orbit of the cislunar case.}
\end{figure}

Next, we consider an NRHO, a periodic orbit near the Libration point in the Earth-Moon three-body system. A 9:2 NRHO, with its orbital parameters listed in Table~\ref{tab:Parameters for the cislunar case}, is selected as the nominal orbit. Such a case is adopted from Ref.~\cite{Boone2022JGCD}. Similar to the Jupiter case (\emph{i.e.}, in Sec.~\ref{Sec:Jupiter Case}), the simplified CRTBP model (\emph{i.e.}, Eq.~\eqref{eq:CRTBP} is employed to describe the dynamics. As illustrated in Fig.~\ref{fig:Nominal orbit of the cislunar case}, the NRHO starts at its apolune and is propagated for a 1.5 orbital period (\emph{i.e.}, $t_0=0$, and $t_f \approx 2.2667$), and finally, it arrives at its perilune. The NRHO has a perilune radius of approximately 1850 km and an apolune radius of approximately 17350 km relative to the Moon. Due to the strong gravitational effects of the Moon around its perilune, the NRHO is particularly sensitive in this region; thus, the linear solution (\emph{i.e.}, the STM) will be insufficient to propagate the uncertainties around the perilune, and a higher-order method is desired \cite{Zhou2023JSR}.

\begin{table}[!h]
    \caption{\label{tab:Parameters for the cislunar case} Parameters for the cislunar case}
    \centering
    \begin{tabular}{lrl}
        \hline\hline
        \multicolumn{2}{l}{Parameter}                     & Value              \\ \hline
        \multicolumn{2}{l}{$\mu$}                         & 0.0121505839705277 \\ \hline
        \multirow{3}{*}{Position vector (nd)} & $x$       & 1.02202815472411   \\
                                              & $y$       & 0                  \\
                                              & $z$       & -0.182101352652963 \\ \hline
        \multirow{3}{*}{Velocity vector (nd)} & $\dot{x}$ & 0                  \\
                                              & $\dot{y}$ & -0.103270818092086 \\
                                              & $\dot{z}$ & 0                  \\ \hline
        \multicolumn{2}{l}{Period (nd)}                   & 1.51119865689808   \\ \hline
        \multicolumn{2}{l}{Initial epoch $t_0$ (nd)}      & 0                  \\ \hline
        \multicolumn{2}{l}{Final epoch $t_f$ (nd)}        & 2.26679798534712   \\
        \hline\hline
    \end{tabular}
\end{table}

\begin{figure}[!h]
    \centering
    \includegraphics[width=0.45\linewidth]{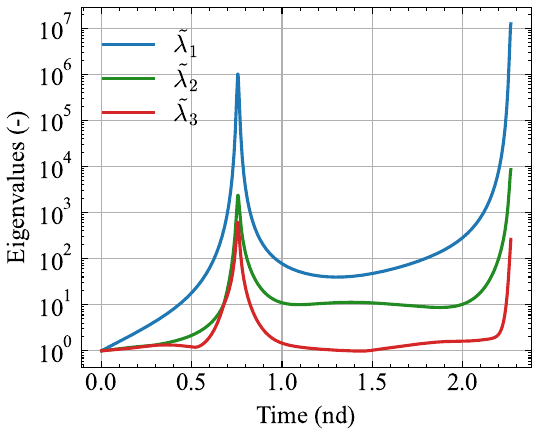}
    \caption{\label{fig:Time histories of the first three eigenvalues of the cislunar case} Time histories of the first three eigenvalues of the cislunar case.}
\end{figure}

\begin{figure}[!h]
    \centering
    \includegraphics[width=0.45\linewidth]{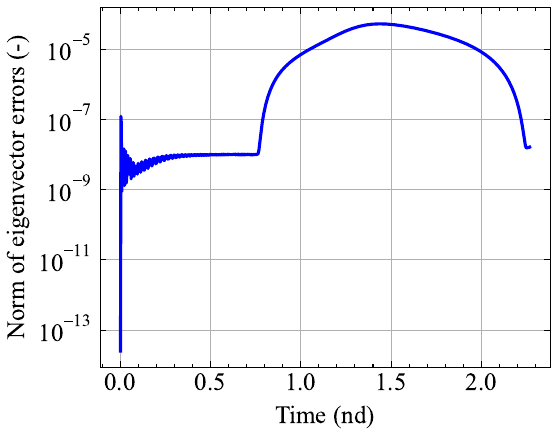}
    \caption{\label{fig:Profiles of the first eigenvalue of the cislunar case} Profiles of the first eigenvalue of the cislunar case.}
\end{figure}

\begin{figure}[!h]
    \centering
    \includegraphics[width=0.45\linewidth]{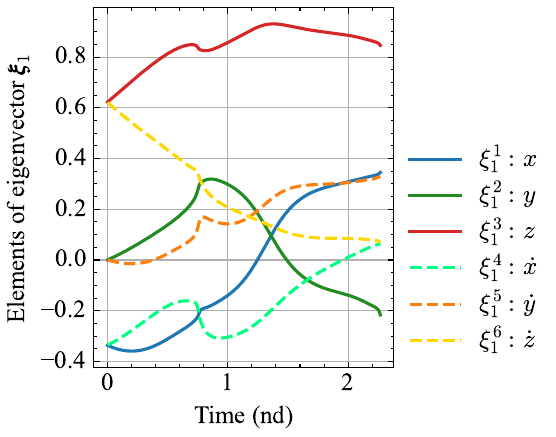}
    \caption{\label{fig:Time histories of the eigenvector elements of the cislunar case} Time histories of the eigenvector elements of the cislunar case.}
\end{figure}

\begin{figure}[!h]
    \centering
    \includegraphics[width=0.45\linewidth]{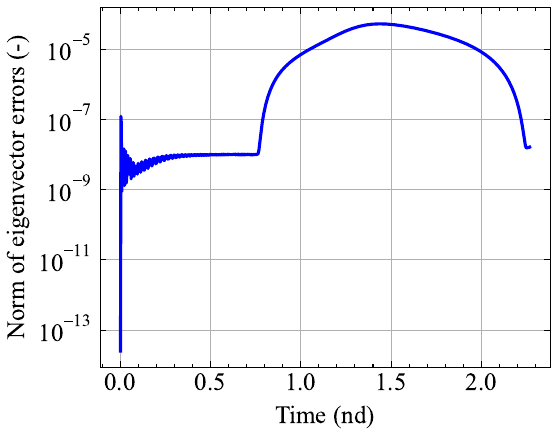}
    \caption{\label{fig:Eigenvector error profile of the cislunar case} Eigenvector error profile of the cislunar case.}
\end{figure}

Figure~\ref{fig:Time histories of the first three eigenvalues of the cislunar case} shows the time histories of the first three eigenvalues of the NRHO over 1.5 periods. The eigenvalues around the perilune points are larger than those around the apolune points, as the dynamics become much more sensitive when the NRHO is closer to the Moon. The largest eigenvalue (\emph{i.e.}, ${\tilde \lambda _1}$) is approximately three orders of magnitude larger than the second largest one (\emph{i.e.}, ${\tilde \lambda _2}$). The profiles of the CGT’s and the TDSTT’s eigenvalues over time are plotted in Fig.~\ref{fig:Profiles of the first eigenvalue of the cislunar case}, with the differences several orders of magnitude smaller the eigenvalues. In addition, the eigenvector (propagated by the TDSTT) and its errors with respect to the CGT's results are shown in Fig.~\ref{fig:Time histories of the eigenvector elements of the cislunar case} and Fig.~\ref{fig:Eigenvector error profile of the cislunar case}, respectively. The elements of the eigenvector change rapidly as time advances. The maximal eigenvector errors are on the level of $10^{-5}$. The results presented in Fig.~\ref{fig:Profiles of the first eigenvalue of the cislunar case} and Fig.~\ref{fig:Eigenvector error profile of the cislunar case} indicate that the TDSTT-propagated eigenvalues and eigenvectors are in good agreement with those obtained from the CGT. Again, the differences shown in Fig.~\ref{fig:Profiles of the first eigenvalue of the cislunar case} and Fig.~\ref{fig:Eigenvector error profile of the cislunar case} are caused by numerical integration errors.

\begin{figure}[!h]
    \centering
    \subfigure[RE of $x(t_f)$]{
        \label{fig:15a}
        \includegraphics[width=0.75\linewidth]{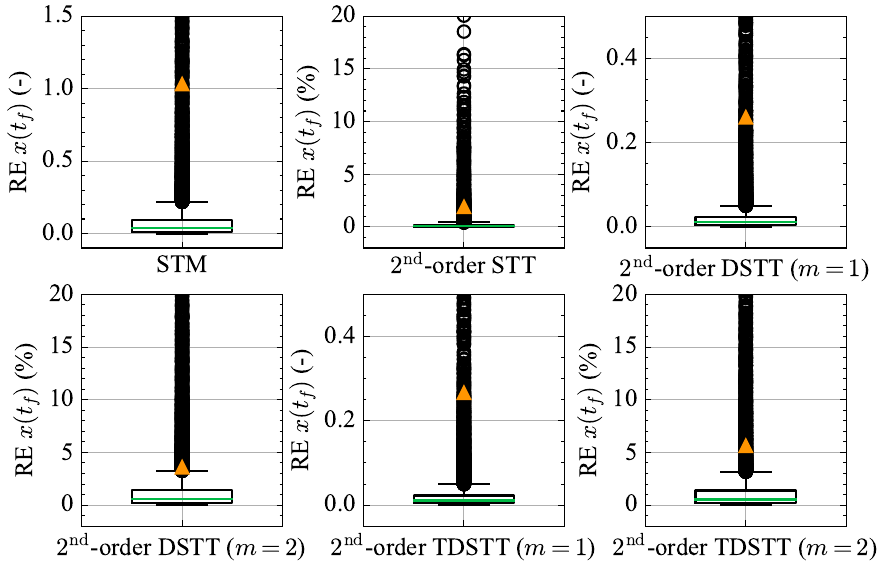}
    }
    \subfigure[RE of $y(t_f)$]{
        \label{fig:15b}
        \includegraphics[width=0.75\linewidth]{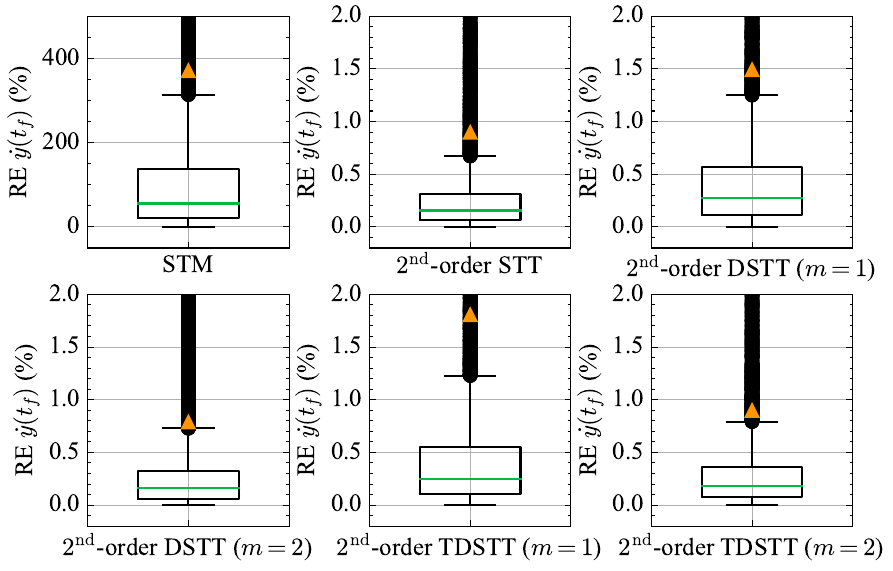}
    }
    \caption{\label{fig:Boxplots of the relative errors in the cislunar case} Boxplots of the relative errors in the cislunar case.}
\end{figure}

Ten thousand MC runs are implemented to compare the accuracy of different methods. The position and velocity STDs for randomly generating initial state deviations are chosen as $2.5 \times 10^{-5}$ per axis and $1 \times 10^{-5}$ per axis, respectively. Figure~\ref{fig:Boxplots of the relative errors in the cislunar case} gives the boxplots of the predicted REs of the linear solution (\emph{i.e.}, STM), the exact high-order solutions (\emph{i.e.}, STTs), DSTTs, and the proposed TDSTTs. In addition, the mean absolute errors (at the final epoch) and the computational time (for integrating the high-order terms) are shown in Table~\ref{tab:Performance comparison of different methods in the cislunar case}. Note that in Table~\ref{tab:Performance comparison of different methods in the cislunar case}, only the mean of the position and velocity error magnitudes (\emph{i.e.}, the norm of the position and velocity error vectors), instead of the MAEs along the \emph{x}-, \emph{y}-, and \emph{z}-axes, are given for simplicity. The computational costs of the DSTT are evaluated based on the direct way (for one propagation from the initial epoch to the predefined final epoch). Recall that the direct way of computing DSTT terms has the limitation of not being able to predict orbital uncertainties at historical points. One can see from Fig.~\ref{fig:Boxplots of the relative errors in the cislunar case} that the proposed TDSTT has box spans close to those of the DSTT. The predicted errors of the TDSTTs are at least one order of magnitude smaller than the errors of the linear method (\emph{i.e.}, STM) and are close to those of the STT and DSTT. Moreover, as shown in Table~\ref{tab:Performance comparison of different methods in the cislunar case}, using two sensitive directions (\emph{i.e.}, $m=2$), the second- and third-order TDSTTs can save 76.28\% and 94.93\% of the computational costs required by the corresponding STTs, respectively. Here, the percentage benefits of the proposed TDSTT over the full STT in terms of computational efficiency are close to those in the Jupiter case (see Table~\ref{tab:Computational costs for computing the high-order terms in the Jupiter case}). The above results show that the TDSTT can obtain nearly the same level of accuracy as the STTs and DSTTs while requiring significantly lower computational costs to compute the high-order terms.

\begin{table}[!h]
    \caption{\label{tab:Performance comparison of different methods in the cislunar case} Performance comparison of different methods in the cislunar case}
    \centering
    \begin{tabular}{llccc}
        \hline\hline
        Order & Method & Computational cost (s) & Position (nd) & Velocity (nd) \\ \hline
        1 & STM & 0.0535 & 2.2413$\times 10^{-6}$ & 4.4198$\times 10^{-4}$ \\ \hline
        \multirow{5}{*}{2} & STT & 12.1099 & 4.1151$\times 10^{-8}$   & 1.5052$\times 10^{-5}$ \\
        & DSTT ($m=1$) & 0.8238 & 1.0925$\times 10^{-7}$ & 1.8788$\times 10^{-5}$ \\
        & TDSTT ($m=1$) & 0.8983 & 3.5805$\times 10^{-7}$ & 3.6145$\times 10^{-5}$ \\
        & DSTT ($m=2$) & 2.5924 & 5.0894$\times 10^{-8}$ & 1.6184$\times 10^{-5}$ \\
        & TDSTT ($m=2$) & 2.8717 & 2.9868$\times 10^{-7}$ & 3.4681$\times 10^{-5}$ \\ \hline
        \multirow{5}{*}{3} & STT & 874.4342 & 2.8846$\times 10^{-8}$ & 1.0238$\times 10^{-5}$ \\
        & DSTT ($m=1$) & 6.7768 & 9.4886$\times 10^{-8}$ & 9.5161$\times 10^{-6}$ \\
        & TDSTT ($m=1$) & 7.6827 & 3.5716$\times 10^{-7}$ & 3.5662$\times 10^{-5}$ \\
        & DSTT ($m=2$) & 41.3918 & 3.0151$\times 10^{-8}$ & 5.9254$\times 10^{-6}$ \\
        & TDSTT ($m=2$) & 44.2694 & 2.9579$\times 10^{-7}$ & 3.1235$\times 10^{-5}$ \\
        \hline\hline
    \end{tabular}
\end{table}

\section{Conclusion}
\label{Sec:Conclusion}
This paper proposed a time-varying directional state transition tensor (TDSTT) for orbital uncertainty propagation, which is an approximation of the traditional state transition tensor (STT). The TDSTT is based on the previous directional STT (DSTT) framework. As TDSTT integrates the eigenvalue-eigenvector pairs and the high-order Taylor series expansions simultaneously, one can employ it to perform analysis at any epoch rather than the predefined final epoch only like the DSTT does. In contrast to the STT, which has an exponential growth in computational burden, the algorithm complexity of the TDSTT grows polynomially with the order. The TDSTT suffers a slight accuracy loss compared to STT and DSTT; however, it can be 60\% faster than the STT at second order and 94\% faster at third order. Additionally, the TDSTT can be hundreds of times faster than the DSTT when used to predict the orbital uncertainties at historical epochs and analyze their evolutions.

\section*{Acknowledgments}
This work was supported by the National Natural Science Foundation of China (No. 12150008, No. 62394353) and the Beijing Institute of Technology Research Fund Program for Innovative Talents (No. 2022CX01008). Xingyu Zhou is grateful for the financial support provided by the China Scholarship Council (Grant No.202406030186).

\bibliography{sample}

\end{document}